\documentclass[11pt,leqno]{article}

\usepackage{amsmath,amssymb,amsfonts,amsthm,mathrsfs,verbatim} 
\usepackage[noend]{algpseudocode}
\usepackage{setspace}
\usepackage{authblk}
\usepackage{caption}
\usepackage{geometry}
\usepackage{graphics} 
\usepackage{graphicx}  
\usepackage{hyperref}
\usepackage{makeidx}
\usepackage{color}
\usepackage{bm}
\usepackage{natbib}

\newtheorem{theorem}{Theorem}[section]
\newtheorem{proposition}[theorem]{Proposition}

\newtheorem{lemma}[theorem]{Lemma}
\theoremstyle{definition}
\newtheorem{remark}[theorem]{Remark}

\newtheorem{algorithm}[theorem]{Algorithm}

\def\R{{\mathbb R}}

\def\talpha{\widetilde{\alpha}}
\def\tbeta{\widetilde{\beta}}
\def\tF{ \widetilde{F}}
\def\btheta{ {\bm \theta} }
\def\balpha{ {\bm \alpha} }
\def\bbeta{ {\bm \beta} }
\def\bkappa{ {\bm \kappa} }
\def\bzeta{ {\bm \zeta} }
\def\bgamma{ {\bm \gamma} }
\def\bxi{ {\bm \xi} }
\def\wkappa{ \widetilde{\kappa} }
\def\wzeta{ \widetilde{\zeta} }
\def\walpha{ \widetilde{\alpha} }
\def\wbeta{ \widetilde{\beta} }
\def\bara{\bar{\bm \alpha}}
\def\barb{\bar{\bm \beta}}

\title{Modelling of Functional Profiles and Explainable Shape Shifts Detection: An Approach Combining the Notion of the Fr\'echet Mean with the Shape Invariant Model}

\date{}
\author[a,c]{Georgios I. Papayiannis}
\author[b,d]{Stelios Psarakis}
\author[b,c]{Athanasios N. Yannacopoulos}

\affil[a]{\footnotesize Section of Mathematics, Department of Naval Sciences, Hellenic Naval Academy, Piraeus, GR 18539}
\affil[b]{\footnotesize Department of Statistics, Athens University of Economics and Business, Athens, GR 10434}
\affil[c]{\footnotesize Stochastic Modelling and Applications Laboratory, Athens University of Economics and Business, Athens, GR 10434}
\affil[d]{\footnotesize Laboratory of Statistical Methodology, Athens University of Economics and Business, Athens, GR 10434}

\begin{document}
\graphicspath{ {figures/} } 
\maketitle

\abstract{A modelling framework suitable for detecting shape shifts in functional profiles combining the notion of Fr\'echet mean and the concept of deformation models is developed and proposed. The generalized mean sense offered by the Fr\'echet mean notion is employed to capture the typical pattern of the profiles under study, while the concept of deformation models, and in particular of the shape invariant model, allows for interpretable parameterizations of profile's deviations from the typical shape. EWMA-type control charts compatible with the functional nature of data and the employed deformation model are built and proposed, exploiting certain shape characteristics of the profiles under study with respect to the generalized mean sense, allowing for the identification of potential shifts concerning the shape and/or the deformation process. Potential shifts in the shape deformation process, are further distinguished to significant shifts with respect to amplitude and/or the phase of the profile under study. The proposed modelling and shift detection framework is implemented to a real world case study, where daily concentration profiles concerning air pollutants from an area in the city of Athens are modelled, while profiles indicating hazardous concentration levels are successfully identified in most of the cases.}\\

\noindent {\bf Keywords:} environmental modelling; EWMA control charts; Fr\'echet mean; functional profiles; shape invariant model; shift detection; statistical modelling; \\

\noindent {\bf MSC:} 62-08, 62P12, 62P30, 62R07, 62R10

\section{Introduction}\label{sec-1}

The task of efficient modelling and the accurate identification of potential shifts for functional profiles has attracted the interest of the statistical community the later years. Such data structures are met more and more in practice in many fields, for instance, in medicine and medical imaging, in electricity management, in water distribution networks, in environmental sciences, etc. The development of more elaborate and complex monitoring schemes is required for the statistical treatment of dynamical and functional data structures in order to reveal and recover important information concerning the special features of the process under study. The resulting functional data (and their increased complexity) is an issue that at a first stage can be handled either by the application of non-linear models or by non-parametric or semi-parametric approaches (e.g. kernel-based estimators, wavelets, etc.). However, as far as statistical control is concerned, for these functional models the classical control tools may not be appropriate; partly because of their functional nature and partly because the output of such models may not be appropriately accommodated in a vector space framework. This fact imposes the need for development of new monitoring mechanisms compatible with the functional data setting. Particularly, in the field of functional profiles monitoring, several approaches have been proposed lately, and to mention just a few, non-parametric regression or wavelets approaches \cite{chicken2009statistical, qiu2010nonparametric, mcginnity2015nonparametric}, interpolation schemes \cite{moguerza2007monitoring, fasso2016functional}, PCA and functional PCA-based methods \cite{shiau2009monitoring, yu2012outlier, paynabar2016change}, functional regression schemes \cite{fasso2016functional, centofanti2020functional, flores2020constructing}, approaches exploiting the intristic geometry of appropriate statistical manifolds or relying on the notion of statistical depth \cite{harris2020elastic, zhao2020intrinsic} and others. The majority of the aforementioned approaches attempt through modifications in the modelling procedure, but still relying on the current setting of process control theory and practice, to extent the current monitoring tools to better treat the new modeling setup, accommodating certain of the salient features of the data in question (such as functional dependence, etc). 

A crucial issue in studying functional profiles from the monitoring perspective, concerns the actual definition of the typical or the so called {\it in control} (IC) behaviour. Standard approaches assume that the under study data can be represented as points of a finite dimensional Euclidean space (e.g. of $\R^d$) of suitable dimensionality related to the number of features under consideration, possibly carrying a correlation structure or displaying variability which is modelled essentially under the assumption of a probability law similar or sufficiently close to a Gaussian one. On the context of functional data, the above setting is not appropriate for a number of reasons. To name just a few:
\begin{itemize}
\item[(a)] The finite dimensionality assumption is no longer valid as the data in question are infinite dimensional, e.g. curves, shapes, surfaces. 

\item[(b)] The dependence structure displayed by data may not be sufficiently modelled within the normality assumption since more complex type of dependencies may be displayed which cannot be approximated by the Gaussian model.

\item[(c)] The observed data may no longer be understood as elements of a vector space but as elements of a space with nonlinear or convex structure (e.g. elements of a general metric space like covariance matrices, curves of particular form, etc).
\end{itemize}

Recently, in \cite{cano2015using} a more appropriate framework for dealing with functional data was proposed, employing tools from the statistical shape theory (see e.g. \cite{dryden2016statistical, small2012statistical}). The current work emphasizes on the appropriate definition and estimation of the typical behaviour of the objects under study, through a more general notion of mean, applicable for metric space valued data, the  Fr\'echet mean \cite{frechet1948elements, le2000frechet}, combined with the framework of deformation models. The approach that is proposed and developed is quite general and can be applied to any functional object (e.g. curves, surfaces, etc.). However, in order to properly motivate the proposed methodology, the case of profiles represented by curves is studied in this work. Attempting to provide a more concrete framework, the case of the shape invariant model (SIM) is considered, a functional modelling approach which is widely applicable to various contexts. Nevertheless, the discussed framework is extendable to any other deformation model choice beyond SIM.

The purpose of this work is twofold. First, a modelling framework for representing functional profiles as deformations of a reference typical profile under the shape invariant model parameterization is proposed. The typical profile is modelled as the Fr\'echet mean of the sample of the provided IC profiles, while the IC distributions of the deformation features according to the SIM parameterization are induced through a registration step of each IC profile to the typical one. Therefore, at this stage the typical IC pattern is recovered while acceptable deformations (or deviations) from the typical pattern are expressed in terms of the obtained distributions of the deformation parameters the SIM incorporates. Secondly, a profile monitoring procedure is built upon these characteristics, to detect potential shifts in newly sampled profiles. In particular, employing the adopted deformation modelling approach and the notion of Fr\'echet mean, a shift detection  scheme is developed on the rationale of an exponentially weighted moving average (EWMA) monitoring procedure that allows for: (a) the detection of potential shifts in the underlying shape of the profile under study and (b) the detection of potential shifts in the deformation process of the profile compared to the typical pattern. The latter step, allows for further investigation of the shift (if occurred) and categorizing it in terms of shifts in certain characteristics related to amplitude or phase. Both steps of the scheme rely on the relevant deviance processes for the underlying shape and SIM deformations, respectively. Note that the control bounds are determined through the quantiles of the induced empirical IC distributions of each attribute.

The paper is organized in the following way. In Section \ref{sec-2} some important preliminaries concerning the notion of the Fr\'echet mean and the framework of deformation models are presented accompanied by a brief literature overview. In Section \ref{sec-3}, our modelling setup combining the notion of the Fr\'echet mean and the shape invariant model is presented accompanied by appropriate numerical approximation schemes for estimating the mean pattern (i.e. the reference profile or the Fr\'echet mean) and, registering a sampled profile to the reference one according to the shape invariant deformation model. Next, in Section \ref{sec-4}, the two-step shape shift detection scheme is proposed, by appropriately extending the EWMA scheme framework to the case of non-vector spaces. At the end of the same Section, the proposed approach is implemented to a real world example where ambient air quality profiles in an area of the city of Athens are modelled, and applying the proposed shift detection method, out of control behaviours are successfully identified in most of the cases.

\section{Literature Overview \& Preliminaries}\label{sec-2}

Here the notion of the Fr\'echet mean and the framework of deformation models are  briefly introduced. These are the fundamental concepts upon which the present work on functional profiles modelling and monitoring relies.

\subsection{Fr\'echet mean: a generalized version of the mean}

Functional objects like profiles, curves, surfaces, etc. are not objects that necessarily belong to a vector space. For instance, if a profile is subject to some distortion, it is not necessary that this output can be represented under the application of linear operations on the possible profiles of the space under study. In fact, the representation of the distorted profile is not guaranteed by a vector space, even if it is known that the distortion is linear with respect to the amplitude and phase, since the aggregate distortion would be nonlinear with respect to the profile itself (see for example the effect of the shape invariant model which is examined in this work). As such, the typical notion of mean is not applicable in this case since there is not the typical vector space setting. This motivates the need for a notion of the mean that does not rely on linearity. Such a generalized notion of the mean which is widely applicable is offered by the Fr\'echet mean \cite{frechet1948elements, zemel2019frechet}, which for a sample of random elements $\{x_1, \ldots, x_n\} \subset \mathcal{M}$ is defined as
\begin{equation}\label{fmdef}
	x_{F} := \arg\min_{z \in \mathcal{M}} \Psi(z)
\end{equation}
where $\Psi(z) = \sum_{j=1}^n w_j d^2(z, x_j)$ is the generalized Fr\'echet function of the sample, $d$ is a suitable metric for the space $\mathcal{M}$ and $\{w_1, \ldots, w_n\}$ is a suitable set of weights for the data (the case $w_{i}=1/n$ corresponds to the standard definition of the Fr\'echet mean).  The minimum value of $\Psi$ (achieved at $x_{F}$) corresponds to the variance of the sample. The Fr\'echet mean is a variational concept that  can be obtained using techniques from the calculus of variations and optimization (see e.g. \cite{ky2020variational}) and its theoretical properties are well documented (see e.g. \cite{le2000frechet, afsari2011riemannian, arnaudon2013medians, petersen2019frechet}). It has been recently used by the statistics community as an important tool in the study of data which cannot be conveniently described as elements of vector spaces (see e.g. \cite{dubey2019frechet}, \cite{izem2007analysis}, \cite{jung2012analysis}, etc). In the monitoring framework presented in this work, the Fr\'echet mean is employed to capture the typical pattern or mean behaviour of an IC sample of functional profiles. However, the performance of this approach significantly depends on the suitability of the deformation model that is chosen to model the variations around the typical pattern (in the case that a semi-parametric approach is adopted) and of course the metric sense under which the operations are derived.

\subsection{The framework of deformation models}

A very natural setting for the study of functional profiles is that of deformation models.  Deformation models produce metric space valued random elements which belong to some set $\mathcal{M}=\{ f \circ T \,\, : \,\, T \in V \}$ where $f$ is a deterministic function  (often to be specified) characterizing the typical (average) shape, and $T(\cdot)=T(\cdot, \omega)$ is a random deformation typically chosen from a vector space $V$ so that the composition $f \circ T(\cdot, \omega)$ generates a random element from $\mathcal{M}$, satisfying certain qualitative features. Hence, each random element is parameterized in terms of the realization of $T \in V$. For example, $f$ can be the curve modeling the distribution of some pollutant over the day, accounting for patterns like the mean variability of traffic or the mean variability of temperature, while $T$ accounts for specific features that may randomly happen on a particular day and may not repeat. Note that in many cases of interest $\mathcal{M}$ is not endowed with a linear structure (e.g. in the case of $s$-shaped curves) but is rather a subset of a more general metric space. For instance, consider curves of a specific shape as those appear in medicine (e.g. intra-day blood pressure curves) or those describing the intra-day demand of electric energy in a city or daily consumption patterns over a water distribution network. Such objects do not necessarily belong to a vector space, since summing two sampled objects does not necessarily leads to an object of the underlying space. This type of statistical modelling has found several applications in practice so far, e.g., in image and signal processing \cite{bigot2009statistical, bigot2011consistency}, in analysing point processes \cite{panaretos2016amplitude}, in medicine \cite{papayiannis2018functional}, in electric energy prediction \cite{papayiannis2020energies}, etc. Restricting attention to the case where $f : I \to \R$ is a function from some interval of $I$ ($I=[0,1]$ without loss of generality) is a suitable choice for the study of nonlinear profiles. A more generic form of a  deformation model has been studied in \cite{panaretos2016amplitude}, modeling separately amplitude and phase deformation characteristics. Under the perspective of the general {\it amplitude-phase} deformation model, the shape relation between two curves $f$ and $g$ is expressed as
\begin{equation}\label{np-smod}
	f(t) = A \left( g(\Phi^{-1}(t)) \right) + \epsilon(t) 
\end{equation}
where $\Phi : I \to I$ for $I\subset \R$ represents the {\it phase deformation process} (its inverse $\Phi^{-1}$ is often referred to as the {\it time-warping function} according to the terminology of \cite{wang1997alignment}) and $A :  L^2(I) \to L^2(I)$\footnote{\small $L^2(I)$: the Lebesgue space of square integrable functions with domain $I$} represents the {\it amplitude deformation process}, while $\epsilon(t)$ denotes the relevant error term, i.e. the aspects of curve $f$ that cannot be efficiently captured through a deformation of $g$ under the particular shape deformation modeling approach. Parametric forms of $\Phi$ and $A$ are conceivable  and practical and  hence often used in practice (see e.g. \cite{kneip1995model, wang1997alignment, gervini2004self}). Considering affine parametric forms for both phase and amplitude deformation process, i.e. $A(z) = \alpha z + b$ and $\Phi(s) = \kappa s + \zeta$, reduces the amplitude-phase deformation model to the well celebrated shape invariant model on which the presented approach relies on.

\section{A Functional Modelling Approach Combining the Shape Inva\-riant Model Framework and the Notion of the Fr\'echet Mean}\label{sec-3}

In this section  a modelling framework for functional profiles is provided which allows for the parameterization of an individual's profile deviance with respect to the typical pattern. In particular, special attention is given to the shape invariant deformation model under which linear deformations are assumed with respect to the amplitude and phase characteristics separately (but still nonlinear with respect to the reference shape). Relying on this deformation model, the Fr\'echet mean of a set of profiles is approximated by appropriately formulating the Fr\'echet variance determination problem and proposing a numerical approximation scheme for this purpose. Moreover, the registration problem of a sampled profile to the reference shape (Fr\'echet mean) is considered, providing the related well posedness results.

\subsection{A brief presentation of the shape invariant deformation model}\label{sec-3.1}

\begin{figure}[ht!]
	\centering
	\includegraphics[width=5.5in]{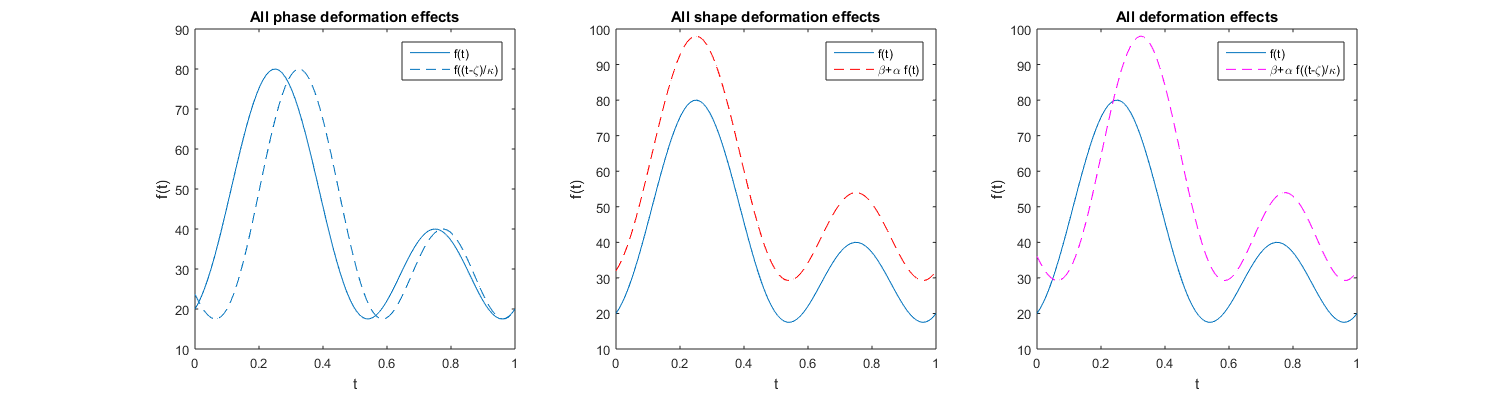}
	\caption{Shape-invariant model parameterization effects for phase deformations (left), shape deformations (center) and both types of deformations (right).}\label{fig-FM-deformations}
\end{figure}

The shape invariant model has been repeatedly discussed and appreciated in the functional modelling literature (see e.g. \cite{kneip1995model, gervini2004self, beath2007infant, bigot2013minimax}) due to its applicability in various different contexts and the interpretability of its deformation parameters. Under this setting, an object $f_j$ is modeled as a distortion of the typical profile $f_0$ through the relation
\begin{equation}\label{model-sim}
	f_j(t) = \beta_j + \alpha_j f_0 \left(\frac{t - \zeta_j}{\kappa_j} \right) + \epsilon_j(t), \,\,\, t\in I
\end{equation}
which can be realized as a special case of the general amplitude-phase deformation model (see e.g. \cite{panaretos2016amplitude}) considering $A_j(z) = \alpha_j z + \beta_j$ and $\Phi_j(s) = \kappa_j s + \zeta_j$ with $(\alpha_j, \beta_j)$ denoting the deformation parameters related to the scale and location of the amplitude deformation process, while $(\kappa_j, \zeta_j)$ denoting the parameters affecting the scale and location of the phase deformation process (please see Figure \ref{fig-FM-deformations} for an illustration of the deformation model's characteristics). In this perspective, the space of profiles $\mathcal{M}$ is identified by the following subset of $L^2$
\begin{equation}\label{Mspace}
\mathcal{M} := \left\{ f\in L^2 \,\,\, :\,\,\, \exists\,\,\, (\alpha, \beta, \kappa, \zeta)' \in \Theta \subset \R^4 \,\,\, \mbox{ where}\,\,\, f(t) = \beta + \alpha f_0\left( \frac{t-\zeta}{\kappa} \right), \,\,\, t\in I\right\}
\end{equation}
with the set $\Theta$ including the feasible values for the deformation parameters, leading to a metric space setting which does not carry the linear structure of a vector space.

\subsection{Determination of the typical profile under the shape invariant model}\label{sec-3.2}

Assume that there is a training set of profiles $\mathcal{F} := \{ f_1, f_2, ..., f_n\}$, and that each profile $f_j$ can be sufficiently modeled as a deformation of the (unknown yet) mean profile $f_{0}$, as represented in \eqref{model-sim}, with deformation parameters $\theta_j = (\alpha_j, \beta_, \kappa_j, \zeta_j)'$. Amplitude distortions of $f_j$ comparing to the reference (Fr\'echet mean) profile $f_{0}$ are captured by the deformation function $A(t; \gamma_j) = \beta_j + \alpha_j f_0(t)$ with $\gamma_j := (\beta_j, \alpha_j)'$ denoting the amplitude deformation parameters, while  phase distortions are captured by the function $\Phi(t; \xi_j) = \zeta_j + \kappa_j t$ with $\xi_j = (\zeta_j, \kappa_j)'$ denoting the relevant phase deformation parameters. These parameterized effects cannot be specified unless the Fr\'echet mean is estimated. Following the current modeling approach, a semi-parametric expression of the mean is obtained through averaging. Using the model \eqref{model-sim}, one obtains a motivation for the definition of Fr\'echet mean using the functional data $f_j$. In particular, one may express
\begin{equation}\label{sim-fm}
	f_{0}(t) = \hat{f}_{0,j}(t; \theta_j) + \eta_j(t)  = \alpha_j^{-1} \left( f_j( \kappa_j t + \zeta_j ) - \beta_j \right) + \eta_j(t)
\end{equation}
where $\eta_j(\cdot)$ denotes the distorted error process related to the estimation error relying on the observation from $f_j$. The mean profile $f_{0}$ must be chosen in such a manner that satisfies the barycenter property, i.e. being simultaneously so close and so far from all the profiles contained in the set $\mathcal{F}$. A standard requirement that has to be met is that the average of the residuals $\eta_j$ among all elements, for each time instant $t\in I$, must be zero, i.e. $\frac{1}{n}\sum_{j=1}^{n}\eta_j(t)=0$. As a result, by properly averaging every possible model of $f_{0}$ as a deformation of each $f_j \in \mathcal{F}$, the following semi-parametric expression is obtained
\begin{equation}\label{fm-sim}
	\hat{f}_{0}(t; \btheta) = \frac{1}{n}\sum_{j=1}^n \hat{f}_{0,j}(t; \theta_j) = \frac{1}{n} \sum_{j=1}^n \alpha_j^{-1} \left( f_j(\kappa_j t  + \zeta_j) - \beta_j \right)
\end{equation}
where $\btheta = (\balpha', \bbeta', \bkappa', \bzeta')'$ with $\balpha = (\alpha_1, ..., \alpha_n)'$, $\bbeta = (\beta_1,...,\beta_n)'$, $\bkappa = (\kappa_1,...,\kappa_n)'$ and $\bzeta = (\zeta_1, ..., \zeta_n)'$. Clearly, what is represented in \eqref{sim-fm} is not the pure Fr\'echet mean, but rather its best approximation under the shape invariant model. The optimality criterion which is employed for the selection of $\btheta$, depends on the metric sense $d_M$ under which the mean element is derived. Recalling the Fr\'echet function defined in \eqref{fmdef}, the optimal $\btheta_* \in \Theta$, where $\Theta$ denotes the feasible set of the deformation parameters, is obtained as the solution to the minimization problem
\begin{equation}\label{opt-2}
	\min_{\btheta \in \Theta } V(\btheta) := \min_{\btheta \in \Theta } \frac{1}{n} \sum_{j=1}^n d^2_M( \hat{f}_{0}(\btheta), f_j )
\end{equation}
leading to the determination of the Fr\'echet mean under the current deformation family of models. In general, the required centrality properties for the amplitude and phase deformation functions are 
\begin{itemize}
\item $A_1 \circ A_2 \circ \cdots \circ A_n = Id$  
\item $\Phi_1 \circ \Phi_2 \circ \cdots \circ \Phi_n = Id$.
\end{itemize}
Specifying the conditions for the case of the shape invariant model, are equivalently stated by the set of constraints for the related deformation parameters
\begin{equation}\label{c-cons}
	\prod_{j=1}^n \alpha_j = 1, \,\,\,\, \sum_{j=1}^n \beta_j = 0, \,\,\,\, \prod_{j=1}^n \kappa_j=1, \,\,\,\, \sum_{j=1}^n \zeta_j = 0
\end{equation}
with $\alpha_j, \kappa_j$ being strictly positive and upper bounded by some $\alpha_{\max}, \kappa_{\max}$, while $\beta_j$ and $\zeta_j$ are constrained to some appropriate intervals $[\beta_{\min}, \beta_{\max}]$ and $[\zeta_{\min}, \zeta_{\max}]$, respectively (for all $j=1,2,...,n$). These properties ensure that the total deformation effect sums up to zero, and that the derived minimizer from \eqref{opt-2} represents the optimal reference profile since, all elements in the training set are aligned with respect to this profile, either in terms of amplitude or phase distortions. To make more clear this conditions, consider for instance the simple case of two profiles. Under the aforementioned centrality conditions, the aggregate amplitude deformation effect is required to satisfy
\begin{eqnarray*}
	(A_1 \circ A_2) ( g ) = g \,\,\, \iff \,\,\, \beta_1 + \beta_2 + \alpha_1 \alpha_2 g = g, \,\,\, \forall g \in L^2
\end{eqnarray*}	
resulting to $\beta_1 + \beta_2 = 0$ and $\alpha_1 \alpha_2 = 1$, while concerning the aggregate phase deformation effect it is required that
\begin{eqnarray*}
	(\Phi_1 \circ \Phi_2) ( t ) = t \,\,\, \iff \,\,\, \zeta_1 + \zeta_2 + \kappa_1 \kappa_2 t = t, \,\,\, \forall t \in I
\end{eqnarray*}	
resulting to $\zeta_1 + \zeta_2 = 0$ and $\kappa_1 \kappa_2 = 1$. Then, the obtained mean element provides a comparison pattern with respect to which, all individual elements in the training set are aligned to in terms of both phase and amplitude.

\begin{remark}
Note that a potential minimizer $\btheta_*$ of \eqref{opt-2} contains the optimal deformation parameters concerning the approximation of the Fr\'echet mean of the set $\mathcal{F}$ under the SIM approach. For a specific profile $f_j \in \mathcal{F}$ the related deformation parameters $\theta_{*,j}$ contained in the vector $\btheta_*$ does not necessarily coincide with the parameter choices that best represent the current profile as a SIM deformation with respect to the mean shape. This happens since optimal solution of \eqref{opt-2} is subject to the set of constraints stated in \eqref{c-cons} which are much more restrictive than the ones that are required for registering some specific $f_j$ to the mean shape (this matter is discussed separately in the following sections). Clearly, $\btheta_*$ is the choice of parameters indicating the optimal position (or tuning) of the semi-parametric mean shape model stated in \eqref{fm-sim} with respect to all elements in the set $\mathcal{F}$. Clearly, this is not necessarily the same with collecting the optimal deformations of each $f_j \in \mathcal{F}$ with respect to the (estimated) mean shape $\hat{f}_0(\btheta)$.
\end{remark}

The problem of estimating the mean profile (typical shape) under the SIM framework and employing the $L^2$ metric sense (i.e. $d_M(f,g) = \| f-g \|_2$ for any $f,g \in \mathcal{M}$) is now considered. The resulting Fr\'echet function minimization problem is characterized by the objective function
\begin{equation}\label{Fobj}
	V(\btheta) := \frac{1}{n}\sum_{j=1}^n \langle f_j - \hat{f}_0(\btheta), f_j - \hat{f}_0(\btheta) \rangle 
\end{equation}
where $\langle f, g \rangle := \int_I f(s)g(s)ds$ denotes the inner-product in the $L^2(I)$ for any $f,g \in L^2(I)$ and $\hat{f}_0(\btheta)$ as stated in \eqref{fm-sim}. Following the discussion in the previous section, the problem of minimizing \eqref{Fobj} over the set $\Theta$, denoting the feasible set for $\btheta$ satisfying all the constraints stated in \eqref{c-cons}, is studied. In order to guarantee the uniqueness of the solution (existence is guaranteed by the continuity with respect to $\btheta$), appropriate regularizations terms are necessary. For this reason, the quadratic regularization term $R_{\rho}(\btheta) = \frac{1}{2\rho}\|\btheta\|_2^2$ is introduced, driven by the sensitivity parameter $\rho > 0$. This regularization transforms, for a certain value of $\rho$ and below, the problem to a strictly convex one. Besides the convexity-correction of the problem, the regularization term enhances the stability of the problem, not allowing to each deformation parameter exceeding a certain threshold value, which in combination with the constraints provided by the set $\Theta$, reduces the effect of potential extreme profiles in $\mathcal{F}$ to the estimation of the mean, robustifying the estimation procedure. Modifying the problem in this way, when more than one minimizers exist, a selection criterion among the potential solutions is introduced, setting a preference order that promotes the ones that cause less dramatic deformations with respect to the original profiles in $\mathcal{F}$. The related result follows (for the relevant proof please see Appendix \ref{App-A-1}).

\begin{proposition}\label{prop-1}
There exists $\rho_0>0$ for which the regularized Fr\'echet variance determination problem 
\begin{equation}\label{reg-ff}
	\min_{ \btheta \in \Theta} \left\{V(\btheta) + \frac{1}{2\rho} \| \btheta \|_2^2 \right\}
\end{equation}
admits unique solution for any $\rho \in (0, \rho_0]$.
\end{proposition}

\subsection{A numerical spitting approach for the approximation of the mean profile}\label{sec-3.3}

Next, the development of a numerical approximation scheme for estimating the mean profile under SIM framework is considered, i.e. approximating the optimal solution of problem \eqref{opt-2}. First, the dependence of function $V$ is distinguished to the amplitude deformation parameters $\bgamma = (\balpha', \bbeta')'$ and to the phase deformation parameters $\bxi = (\bkappa', \bzeta')'$. Splitting the parameters into two distinct groups is a key step towards the development of an efficient numerical approximation scheme. Ignoring at first the centrality constraints \eqref{c-cons} and under appropriate transformations of the parameters contained in $\bgamma$\footnote{considering $\bgamma := (\bar{\balpha}', \bar{\bbeta}')'$ with $\bar{\alpha}_j :=  1/\alpha_j$ and $\bar{\beta}_j := \beta_j \bar{\alpha}_j$ for $j=1,2,...,n$}, the function $V$ depends quadraticaly on the amplitude deformation parameters, while as far as the parameter vector $\bxi$ is concerned, the dependence is non-quadratic and importantly non-convex (at least in general cases). Treating directly the problem with respect to all the parameters, results to a non-convex and computationally expensive minimization problem which is difficult to treat. Alternatively, an iterative parameter-splitting minimization scheme is proposed, which treats the problem in two separate stages and takes advantage over the quadratic dependence of $V$ with respect to $\bgamma$ and by convexly approximating the solution with respect to $\bxi$. Tikhonov-type regularizations are considered to both parts of the problem (for different reasons) and through successive projections of the obtained solutions to the constrained sets an optimal solution is reached.  

The space $\Theta$ is decomposed to two components  $\Theta_{\Lambda}$ and $\Theta_{\Pi}$ (satisfying $\Theta = \Theta_{\Lambda} \times \Theta_{\Pi}$) as follows
\begin{eqnarray}
	\Theta_{\Lambda} &=& \left\{ \bgamma = ( \bar{\balpha}', \bar{\bbeta}' )' \in \R^{2n}\,\,\, : \,\,\, \prod_{j=1}^n \bar{\alpha}_j = 1, \,\,\, \sum_{j=1}^n \frac{\bar{\beta}_j}{\bar{\alpha}_j} = 0, \,\,\, \bar{\alpha}_j >0, \,\,\, \forall j \right\}\\ 
	\Theta_{\Pi} &=&  \left\{ \bxi = (\bkappa', \bzeta')' \in \R^{2n} \,\,\, : \,\,\, \prod_{j=1}^n \kappa_j = 1, \,\,\, \sum_{j=1}^n \zeta_j = 0,\,\,\, \zeta_j \in \left[\zeta_{\min}, \zeta_{\max} \right], \,\,\, \forall j  \right\}.
\end{eqnarray}
To clarify the separate dependence of $V$ to the amplitude deformation parameters $\bgamma$ and to the phase deformation parameters $\bxi$, is denoted equivalently from now on the respective Fr\'echet function as $V(\bgamma, \bxi)$. Regularization terms of the form suggested in Proposition \eqref{prop-1} are introduced to the problem separately for $\bgamma$ and $\bxi$. Fixing a choice $\bxi$, the (unconstrained) minimization problem with respect to $\bgamma$ can be solved in closed form, while a regularization term is required for stabilizing the estimation procedure. Then, the solution is projected to its constrained set $\Theta_{\Lambda}$. Subsequently, fixing the obtained $\gamma$, the (unconstrained) minimization problem with respect to $\bxi$ is solved, introducing a regularization term that will guarantee the convexity of the objective function. The obtained solution $\bxi$ is projected to its constraint set $\Theta_{\Pi}$ and the above steps are repeated till convergence. The scheme in algorithmic formulation is described below.

\begin{algorithm}[Iterative splitting-projection scheme for Fr\'echet mean approximation]\label{alg-2} \mbox{ }
	
	\begin{itemize}
		\item[ {\bf 0.} ] Set $k=1$, $\bgamma^{(0)} = (1,1,...,1,0,0,...,0)'$, $\bxi^{(0)} = (1,1,...,1,0,0,...,0)'$ and specify a tolerance level ($\epsilon>0$).
		
		\item[ {\bf 1.} ] Solve the amplitude-unconstrained optimization problem 
		\begin{equation}\label{sub-opt-1}
			\hat{\bgamma} = \arg\min_{ \bgamma \in \R^{2n} } \left\{ V(\bgamma, \bxi^{(k-1)}) + \frac{1}{2\rho} \left( \| \bgamma\|_2^2 \right\} + \|\bxi^{(k-1)}\|_2^2 \right) , \,\,\,\, \rho > 0
		\end{equation}
		and then compute the projection $ \bgamma^{(k)} = Proj_{\Theta_{\Lambda}}(\hat{\bgamma})$
		
		\item[ {\bf 2.} ] Solve the phase-unconstrained optimization problem 
		\begin{equation}\label{sub-opt-2}
			\hat{\bxi} = \arg\min_{ \bxi\in \R^{2n} } \left\{ V(\bgamma^{(k)}, \bxi) + \frac{1}{2\rho} \left( \|\bxi\|_2^2 + \|\bgamma^{(k)}\|_2^2 \right) \right\}, \,\,\,\, \rho > 0 
		\end{equation}
		and then compute the projection $ \bxi^{(k)} = Proj_{\Theta_{\Pi}}(\hat{\bxi}) $
		
		\item[ {\bf 3.} ] If $\| \bgamma^{(k)}-\bgamma^{(k-1)} \|<\epsilon$ and $\| \bxi^{(k)}-\bxi^{(k-1)} \|<\epsilon$ stop and set $(\bgamma_*, \bxi_*) = (\bgamma^{(k)}, \bxi^{(k)})$.  Else set $k = k + 1$ and return to {\it Step 1}.
	\end{itemize}
\end{algorithm} 

The aforementioned alternative projections-splitting scheme converges to a unique solution from standard arguments from the theory of alternating projections. The result is stated in the following proposition (for more details please see Appendix \ref{App-A-2}).

\begin{proposition}\label{prop-2}
	The following hold:
	\begin{itemize}
		\item[a.] Given $\bgamma^{(0)} \in \Theta_{\Lambda}$ and $\bxi^{(0)} \in \Theta_{\Pi}$, there exists $\rho^{*}>0$ for which the optimization problems \eqref{sub-opt-1} and \eqref{sub-opt-2} admit unique solutions for any $\rho \in (0,\rho^*]$.
		
		\item[b.] If $\rho$ belongs to $(0,\rho^*]$, then the iterative spitting-projection numerical scheme stated in Algorithm \ref{alg-2} converges to a unique solution.

	\end{itemize}
\end{proposition}

\begin{remark}
	Note that the solution of Problem \eqref{reg-ff} can be obtained by Algorithm \ref{alg-2} for appropriate choices of the individual regularization parameters $\rho_{\Lambda}$ and $\rho_{\Pi}$.
\end{remark}

\subsection{Profile registration to the mean shape}\label{sec-3.4}	

Next, the registration problem of a profile $f_j$ (not necessarily belonging to the training set $\mathcal{F}$) to the mean profile $f_0$ is considered. This can also be realized as the best approximation problem of a $f_j$ through the SIM deformation of $f_0$ (as estimated through the procedure presented in Section \ref{sec-3.3}). This leads to the profile registration problem
\begin{equation}\label{sim-reg}
	\min_{\theta \in \Theta_R} V_j(\theta) = \min_{\theta \in \Theta_R} \langle f_j - \hat{f}(\theta), f_j - \hat{f}(\theta) \rangle
\end{equation}
where $\hat{f}(\theta) = \beta + \alpha f_0( \kappa^{-1}(t-\zeta))$ and 
\begin{equation}\label{Theta-Reg}
	\Theta_R := \left\{ \theta = (\alpha, \beta, \kappa, \zeta)' \,\, : \,\, 
		\begin{array}{ll}
			\alpha \in (0,\alpha_{\max}], & \beta \in [\beta_{\min}, \beta_{\max}],\\ 
			\kappa \in (0,\kappa_{\max}], & \zeta \in [\zeta_{\min}, \zeta_{\max}] 
		\end{array}	
			\right\} 
\end{equation}	
with the lower and upper bounds of the parameters are subject to the characteristics of the profiles $f_0$ and $f_j$. Problem \eqref{sim-reg} although admits solutions by standard continuity arguments, it is not necessary for the solution to be unique. This problem practically concerns only the phase deformation parameters $(\kappa, \zeta)$, since the same problem is strictly convex (quadratic) with respect to the amplitude deformation parameters $(\alpha, \beta)$. Therefore, a regularization term is needed for the successful treatment of the problem which will importantly play the role of a selection criterion when more than one minimizers exist. In that case, one would like to choose that value for $\theta$ which causes the smallest possible deformation to the reference shape $f_0$ in order to represent $f_j$. In the same spirit with Section \ref{sec-3.3}, quadratic regularization terms are incorporated for the appropriate tuning of the aggregate deformation effect, caused by the choice of an arbitrary $\theta=(\alpha,\beta,\kappa,\zeta)'$, favouring to the minimum possible deformation. Then, it can be shown that for a certain value of the sensitivity parameter $\rho$ and below, the single profile registration problem admits unique solution. However, note that the introduced regularization term is not necessary to involve all the parameters in $\theta$, but only $\kappa,\zeta$, since closed-form solutions for $\alpha, \beta$ can be attained given $\kappa,\zeta$ (please see the relevant discussion in the Appendix \ref{App-A-3}).

\begin{proposition}\label{prop-3}
	Let $f_0 \in L^2(I)$ be a reference shape and $f_j \in L^2(I)$ a profile to be registered to it as a SIM deformation. There exists $\rho_*>0$ for which the single profile registration problem 
	\begin{equation}\label{profile-reg}
		\min_{\theta \in \Theta_R} \left\{ V_j(\theta) + \frac{1}{2\rho}\|\theta\|_2^2 \right\}
	\end{equation}	
	where $V_j$ determined in \eqref{sim-reg}, admits unique solution for any $\rho \in (0, \rho_*]$.
\end{proposition}

For the proof please see Appendix \ref{App-A-3}.

\section{The Exponentially Weighted Fr\'echet Moving Average scheme for functional profiles}\label{sec-4}

Following the characterization and estimation of the typical profile in Section \ref{sec-3}, the construction of an EWMA-type proceedure for detecting potential shifts with respect to the underlying shape and deformations from it, is proposed. The main focus is given on the case of curves representing the evolution of a quantity in a specified time (or space) interval $t\in I$, the pattern of which can be efficiently calibrated by the shape invariant model as discussed above. The rationale behind the developed control chart, is similar to the classical EWMA charts procedure either for monitoring the mean or variability process of the statistical quantities under study (see e.g. \cite{montgomery2009statistical}).

The novelty of the proposed shift detection scheme relies on the fact that a generalized mean sense is incorporated directly in the chart element estimation procedure, providing an attractive framework for analyzing functional data parameterizing certain characteristics of them. In what follows, a modification of the EWMA monitoring approach is presented, in which, the mean sense derived by the notion of the Fr\'echet mean is deployed for the development of an exponentially weighted scheme, able to be implemented directly to the metric space of the profiles under study. This modification enables the development of the first branch of the proposed shift detection scheme, which allows for the monitoring of significant changes to the underlying shape of the profile under study. At the second step, a Fr\'echet mean based scheme is built for the surveillance of the shape deformation processes (according to the shape invariant model) which quantify the deviations of the profile under study from the reference pattern. The aforementioned steps provide a two-stage shift detection framework capable of explaining potential shifts of the profile under study and interpret them either as {\it shape shifts} (i.e. significant shift to the underlying shape) or {\it deformation shifts} (i.e. significant shifts to the aggregate deformation effect with respect to the reference shape). In the sequel, the term Exponentially Weighted Fr\'echet Moving Average (EWFMA) is used when referring to the proposed scheme.

\subsection{Shift detections on the underlying shape}\label{sec-4.1}

Given a training set of $n$ in control (IC) profiles $\mathcal{F} = \{ f_1,...,f_n \}$, the typical profile $f_0$ is estimated through the procedure described in Section \ref{sec-3}, under the assumption that any IC element is acceptably deviant from the mean shape $f_0$ with respect to the shape invariant model. Being acceptably deviant means both that (a) the underlying shape of the profile and (b) the respective amplitude-phase deformation process (as captured by the shape invariant model), produce aggregate deviance levels that do not exceed the ones observed from the IC dataset $\mathcal{F}$. Given the estimated IC behaviour $f_0$, for a newly sampled $f_j$, the best approximation according to model \eqref{model-sim} is  obtained through the solution of the related registration problem stated in \eqref{profile-reg}.

At the first stage of the EWFMA scheme, a chart for monitoring the shape deviance process is constructed, i.e. the deviance of the underlying shape for a profile $f_j$ (after removing the amplitude and phase deformation characteristics as estimated from \eqref{profile-reg}) from the reference one. The retrieved underlying shape for the $j$-th profile given the obtained optimal deformation parameters $\theta_j$, is defined as
\begin{equation}\label{shape-j}
	\hat{f}_{0,j} = \alpha_j^{-1} \left( f_j(\kappa_j t +\zeta_j) - \beta_j \right).
\end{equation}
Then, the underlying shape deviance between the $j$-th individual's underlying shape and the typical shape $f_0$ is calculated as
\begin{equation}\label{shape-j-dev}
	D^s_j := d_M^2( \hat{f}_{0,j}, f_0 )
\end{equation}
where $d_M$ denotes the metric sense under which the deviance is determined ($L^2$-distance in this paper). If the underlying shape for a profile is dramatically divergent from the typical one, then the shape invariant model is incapable of modelling the current observation and therefore the examined profile should be considered as an OOC one (i.e. not explainable from the current deformation model). As a result, the following EWMA-type scheme for monitoring the underlying shape deviance process is proposed:
\begin{eqnarray}\label{shape-dev}
	\left\{ 
	\begin{array}{l}
		\widetilde{D}_j^{s} = \lambda D_j^s + (1-\lambda) \widetilde{D}^{s}_{j-1}, \,\,\,\, \lambda \in (0,1], \,\,\,\, j=1,2,... \\
		\widetilde{D}^{s}_{0} = 0
	\end{array}
	\right.
\end{eqnarray}
The bounds beyond which the deviance process provides a shift, can be estimated by the empirical distribution of shape deviances from the IC profiles given certain confidence levels.

\subsection{Shift detections on the deformation features}\label{sec-4.2}

Assuming that no shift in the underlying shape is detected, the profile under study can be described by the shape invariant model, while at the next step, its deformation characteristics (amplitude and phase deformations) are examined. For this task, a separate exponentially weighted chart is developed, relying on the formulation of the shape invariant model combined with the notion of the Fr\'echet mean. The resulting chart concerns the deformations deviance  process, i.e. the total deformation effect performed by both amplitude and phase deformations to the underlying shape. In this view, each profile (since it has been already considered as in control in terms of preserving the typical shape $f_0$) is identified by its modelling counterpart, i.e. $\hat{f}_j = \hat{f}(\theta_j)$ where $\theta_j$ is the minimizer of the related registration problem stated in \eqref{profile-reg} and $\hat{f}(\theta)$ representing the fitted part of $f_j$ from \eqref{model-sim} (without the error term). This second stage of the EWFMA scheme is built directly on the modelled profiles with respect to the Fr\'echet mean sense. Initializing the monitoring procedure with the reference profile $f_0$, each new element of the scheme is naturally derived as the Fr\'echet mean of the profiles $f_0$ and $\hat{f}_j$ with corresponding weights $\lambda$ and $1-\lambda$, leading to the variational problem
\begin{equation}\label{var-problem-profiles}
	\widetilde{f}_j := \arg\min_{g \in \mathcal{M}} \left\{ \lambda d_M^2( g, \hat{f}_j ) + (1-\lambda) d_M^2( g,  \widetilde{f}_{j-1} ) \right\}, \,\,\,\, \widetilde{f}_0 = f_0.
\end{equation}
Since the study of the deformation characteristics with respect to $f_0$ is of particular interest in this work, a reduced version of the variational problem stated above is considered, by substituting $g = \hat{f}(\theta)$ resulting to the simpler fitting problem
\begin{equation}\label{ewma-opt-1}
	\min_{\theta \in \Theta_R} V_{\lambda}(\theta) =  \min_{\theta \in \Theta_R} \left\{ \lambda \| \hat{f}(\theta) - \hat{f}_j \|^2_2 + (1-\lambda) \| \hat{f}(\theta) - \widetilde{f}_{j-1} \|^2_2 \right\}, \,\,\,\, \lambda \in (0,1]
\end{equation}	
where $\Theta_R$ is the set defined in \eqref{Theta-Reg}. For the same reasons discussed in Section \ref{sec-3.4}, the above problem does not necessarily admit unique solution, and a regularized version of the problem similar to the one in Proposition \ref{prop-3} is considered. In a similar manner, the regularization term turns the problem to a strictly convex one with respect to $\theta$ and at the same time introduces a selection criterion when multiple minima occur. The related result follows (for details on the proof please see Appendix \ref{App-B-1}).

\begin{proposition}\label{prop-4}
	Let $f_0\in L^2(I)$ be the mean pattern and $\hat{f}, \widetilde{f} \in L^2(I)$ be two reference profiles. There exists $\widetilde{\rho}_*>0$ for which the chart element estimation problem 
	\begin{equation}\label{cewma-reg}
		\min_{\theta \in \widetilde{\Theta}} \left\{ V_{\lambda}(\theta) + \frac{1}{2\rho}\|\theta\|_2^2 \right\}
	\end{equation}	
	where $V_{\lambda}$ determined in \eqref{ewma-opt-1}, admits unique solution for any $\rho \in (0, \widetilde{\rho}_*]$.	
\end{proposition}

Following the above result, in the general setting the EWFMA chart elements are obtained through the scheme
\begin{eqnarray}\label{ewma-reg}
	\left\{ 
	\begin{array}{l}
		\widetilde{\theta}_j = \arg\min_{\theta \in \widetilde{\Theta}} \left\{ \lambda \| \hat{f}(\theta) - \hat{f}_j \|^2_2 + (1-\lambda) \| \hat{f}(\theta) - \widetilde{f}_{j-1} \|_2^2 + R_{\rho}(\theta) \right\}, \\
		\widetilde{\theta}_0 = (\widetilde{\alpha}_0, \widetilde{\beta}_0, \widetilde{\kappa}_0, \widetilde{\zeta}_0)'=(1,0,1,0)'
	\end{array}\right.
\end{eqnarray}
Determining the optimal parameters $\widetilde{\theta}_j$ is equivallent to the determination of $\widetilde{f}_j$ in problem \eqref{var-problem-profiles} under the shape invariant model framework. Then, the modelled profiles chart elements are obtained in a compatible to the space $\mathcal{M}$ sense, through the notion of the Fr\'echet mean. The relevant chart for monitoring the modelled profiles deviance process (or simply the deformations deviance process) is constructed as 
\begin{eqnarray}\label{deformation-dev}
	\left\{ 
	\begin{array}{l}
		\widetilde{D}_j^{\theta} = \lambda D_j^{\theta} + (1-\lambda) \widetilde{D}^{\theta}_{j-1}, \,\,\,\, \lambda \in (0,1], \,\,\,\, j=1,2,... \\
		\widetilde{D}^{\theta}_{0} = 0
	\end{array}
	\right.
\end{eqnarray}
where $D_j^{\theta} := \| \hat{f}(\theta_j) - f_0\|^2_2$ with control limits of the chart to be specified by the induced empirical distribution from the training set.

\begin{remark}
	One may consider beginning the shift detection procedure by building typical EWMA charts on the obtained deformation parameters and then extending this chart for the deformation processes and the object $f_j$, as well. However, non-linearity of $f_j$  with respect to the parameters (in particular for the shape invariant model with respect to the phase deformation parameters) does not allow for such a generalization. Notice that assuming that the deformation functions $\Lambda, \Pi$ are linear with respect to their parameters, e.g. $\Lambda(z) = \beta + \alpha z$, $\Pi(s)=\zeta + \kappa s$, although one may attempt to monitor the deformation functions separately, e.g. $\widetilde{\Lambda}_j = \lambda \Lambda_j + (1-\lambda) \widetilde{\Lambda}_{j-1}$ and $\widetilde{\Pi}j = \lambda \Pi_j + (1-\lambda) \widetilde{\Pi}_{j-1}$, then the resulting chart for $f_j$ through $\widetilde{f}_j = \lambda \widetilde{\Lambda}_j \circ f_0 \circ \widetilde{\Pi}j + (1-\lambda) \widetilde{f}_{j-1}$ does not coincide with the minimizer of equation \eqref{var-problem-profiles}.
\end{remark}

\subsection{The functional profiles shift detection algorithm}\label{sec-4.3}

\begin{figure}[ht!]
	\centering
 \includegraphics[width=\linewidth]{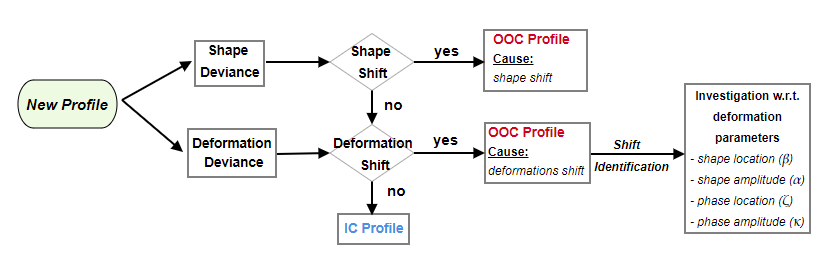}
	\caption{Graphical illustration of the presented scheme profile monitoring process.}\label{ms}
\end{figure}

Combining the two shift detection schemes described in \eqref{shape-dev} and \eqref{deformation-dev} results to an effective tool for both detecting and explaining potential shifts for profiles which can be captured by the parameterization offered by the shape invariant model. At the first step, the observed profile's underlying shape is checked, if it is sufficiently close to the estimated typical pattern ($f_0$). If a shift is detected at this step, then the profile is considered as an out of control (OOC) one, at least under the shape invariant model's explainability. If no shift is detected, then the shape invariant model can be considered as a valid approach and the scheme proceeds with studying the profile's amplitude-phase deformation characteristics. At this step, the second monitoring approach is applied and determines whether the aggregate deformation effect leads to an IC status for the examined profile or not. If no shift is detected, the profile is considered as an IC one, while in the opposite case, a further investigation is performed on which deformation attributes led to the shift by constructing typical EWMA charts on each one of the deformation parameter in $\theta_j$. A graphical representation of the monitoring procedure is illustrated in Figure \ref{ms} while the steps of the monitoring scheme are briefly described in Algorithm \ref{alg-1}.

\begin{algorithm}[The EWFMA shifts detection scheme]\label{alg-1} \mbox{} 
	
	\begin{itemize}	
		\item {\bf Phase I: Mean profile and IC profiles characteristics determination}
		\subitem (a) \emph{Mean pattern estimation}\\
		Given a set of IC profiles $\mathcal{F} = \{ f_1, f_2, ..., f_n \}$ estimate the mean profile $f_{0}$ according to the procedure described in Section \ref{sec-3.2}.
		
		 \subitem (b) \emph{Estimation of the IC profiles deformation characteristics}\\
		 Determine the deformation characteristics of all IC profiles with respect to $f_0$ following the procedure described in Section \ref{sec-3.4}
		
		\item {\bf Phase II: Detection of shape and/or deformation shifts}
		\subitem (a) \emph{Shift detection on the underlying shape}\\
		Given $\lambda>0$, for a newly sampled profile $f_j$ apply the procedure described in \eqref{shape-dev} and determine if the underlying shape of the profile $f_j$ is significantly shifted w.r.t. the reference mean shape $f_0$. If no shift is detected, proceed to Step II(b).
		
		\subitem (b) \emph{Shift detection on the deformation process}\\
		Apply the chart described in \eqref{deformation-dev} to determine if the aggregate amplitude-phase deformation is an IC one. If shift detected, check individual deformation characteristics (deformation parameters) through individual EWMA charts for the parameters obtained by \eqref{ewma-reg}.
	\end{itemize}
\end{algorithm}

\subsection{Analysis of ambient air quality profiles in the area of Athens}\label{sec-4.4}

In this section, the presented methodology is implemented to the analysis of air pollution profiles in an area of the city of Athens. The discussed approach has been also priory tested with success to a number of simulation experiments with synthetic types of data. Since, all the important aspects and capabilities of the method that have been displayed in the synthetic data experiments are fully observed through the air pollution case study without losing any part of the stress assessment of the approach, and taken into account: (a) the better insights provided by a real world example and (b) having in mind to keep the paper coinsize, only the real world example is illustrated. The functional modelling and shift detection approach presented in Section \ref{sec-3} is implemented in studying daily air pollution profiles which measurements have been sampled by atmospheric pollution sensors from the area of Patission Street in Athens. 

The task of monitoring ambient air quality in this area for the time period 2001-2007 under the functional data setting is performed. The available data\footnote{\scriptsize Data are publicly available at:\\ \url{https://ypen.gov.gr/perivallon/poiotita-tis-atmosfairas/dedomena-metriseon-atmosfairikis-rypansis/}} are provided by pollution sensors in the area which have been installed by the Hellenic Ministry of Environment and Energy and consist of hourly measurements (mean values per hour for the duration of a day, i.e. $24$ measurements) of the concentration of four chemical quantities considered as potential pollutants, and specifically, the concentration of: carbon monoxide (CO), nitrogen dioxide (NO$_2$), ozone (O$_3$) and sulfur dioxide (SO$_2$). Certain safety concentration thresholds for the human health concerning the pollutants concentration in the regional atmosphere have been set by the recent directions of the World Health Organization\footnote{\scriptsize \url{https://www.who.int/news-room/fact-sheets/detail/ambient-(outdoor)-air-quality-and-health}} (WHO) (please see Table \ref{tab-4.1}). According to these guidelines, a day in which at least one of the thresholds has been violated is characterized as an OOC day, while if none has been violated is considered as an IC day. This classification strategy is adopted in the experiment for distinguishing between IC and OOC profiles. However, note that these thresholds are constantly revised by WHO with the tendency to become more strict through the years.

\begin{table}[ht!]\small
	\centering
	\begin{tabular}{l|c|c|c}
		\hline\hline
		{\bf Pollutant} & \bf Time Interval & \bf Unit & \bf Maximum Average\\
		&&& \bf Concentration Threshold\\
		\hline
		Carbon Monoxide (CO)      & 1 hr   & mg/m$^3$    &$\leq 35$ \\ 
		& 8 hrs  & mg/m$^3$    &$\leq 10$  \\
		& 24 hrs & mg/m$^3$    &$\leq 4$  \\
		\hline
		Nitrogen Dioxide (NO$_2$) & 1 hr   &$\mu$g/m$^3$ &$\leq 200$ \\ 
		& 24 hrs &$\mu$g/m$^3$ &$\leq 25$ \\
		\hline
		Ozone (O$_3$)             & 8 hrs  &$\mu$g/m$^3$ &$\leq 60$\\
		\hline
		Sulfur Dioxide (SO$_2)$   & 24 hrs &$\mu$g/m$^3$ &$\leq 40$\\
		\hline\hline
	\end{tabular}
	\caption{Safety concentration thresholds of the pollutants under study for human health according to the World Health Organization}\label{tab-4.1}
\end{table}

This threshold-based approach is maybe not the best strategy for the efficient and successful monitoring of ambient air quality, since the actual nature of data is of functional form while the threshold is defined pointwise (with respect to time and space). Under this perspective, it could be possible a pollutant's concentration to be in quite high levels during all the day but never exceeding the nominal threshold value. Therefore such a profile, although not considered as an OOC one, it would provide significant risks for the public health. The discussed functional approach provides an alternative tool for assessing and measuring the status of such phenomena that evolve simultaneously with similar manner (with small fluctuations around a specific standard). Employing the framework of deformation models in the profile modelling task, and in particular exploiting the explanatory capabilities of the shape invariant model, allows for the recovery and quantification of special characteristics that cannot be captured by typical pointwise approaches. To illustrate the capabilities of the discussed method, the recorded profiles from the time period 2001-2004 are used as the training dataset, while as test dataset are used the records from the time period 2005-2007. Note that the focus in the analysis refers to the months October, November and December, since this period is considered as the "peak period" for the concentration of pollutants in the atmosphere (i.e. it is more interesting in terms of identifying the OOC days). Moreover, in these three months no significant differences in the environmental conditions are observed and the median levels of the polluters are quite the same (similar median estimated curves) and as a result, seasonality effect is not an issue. 

\begin{figure}[ht!]
	\centering
	\includegraphics[width=5.5in]{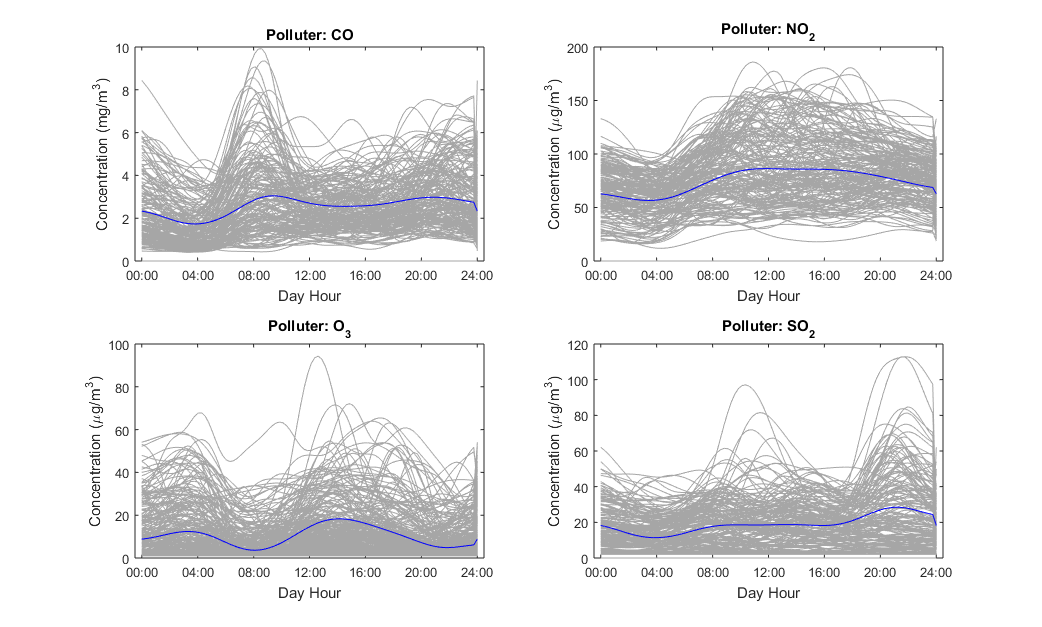}
	\caption{The calculated typical profiles (blue lines) for the intra-day (per hour) concentration of each one of the pollutants (CO, NO$_2$, O$_3$, SO$_2$) in the regional atmosphere.}\label{fig-4.2}
\end{figure}

During the phase I of the monitoring procedure, the IC observations from the time period 2001-2004 are used as the trainset and the typical behaviour for the intra-day concentration of each pollutant (i.e. the typical curve representing the daily evolution of each pollutant) alongside with the respective deformations around the mean patterns and their induced empirical distributions are estimated. A slightly modified version of the shape invariant model is incorporated where the phase scaling parameter $\kappa$ is omitted to simplify the analysis. The deformation characteristics parameterized by the SIM (location, amplitude and phase) allow for a more careful and meaningful representation of the potential deviations from the reference pattern that could possibly affect the pollutant's status at each day. Each one of the considered pollutants display quite different concentration behaviour which enables the assessment of the presented methodology under different conditions and data patterns. In Figure \ref{fig-4.2} are illustrated the IC daily concentration curves of the pollutants under study (recorded in the training set) and their estimated mean patterns (typical concentration profiles). All data patterns seems to be successfully calibrated by the shape invariant model and the calculated Fr\'echet means, representing plausibly the typical behaviours of the pollutants' concentration in the city's regional atmosphere.

\begin{figure}[ht!]
	\centering
	\includegraphics[width=4.2in]{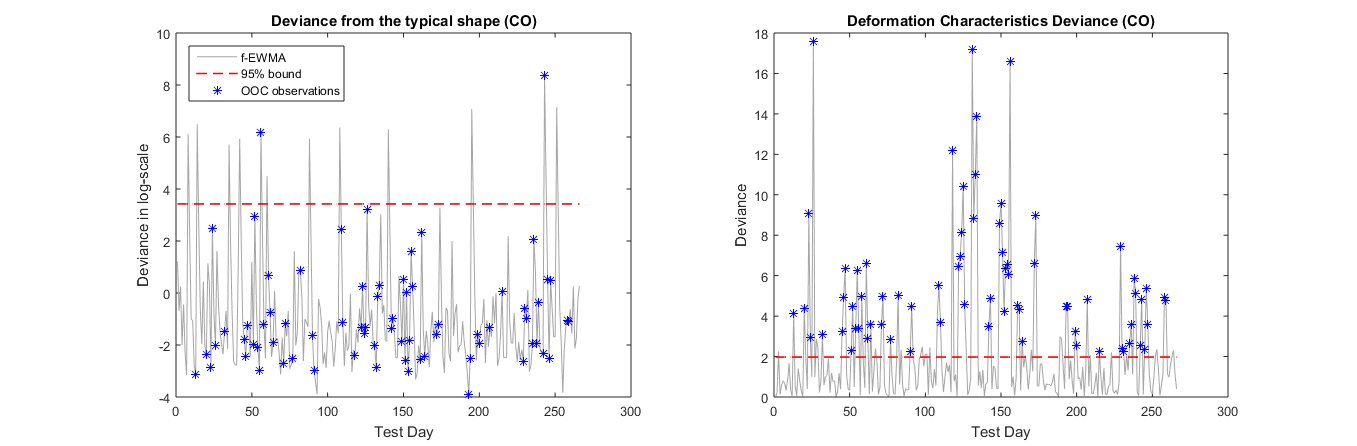}\\
	\includegraphics[width=5.7in]{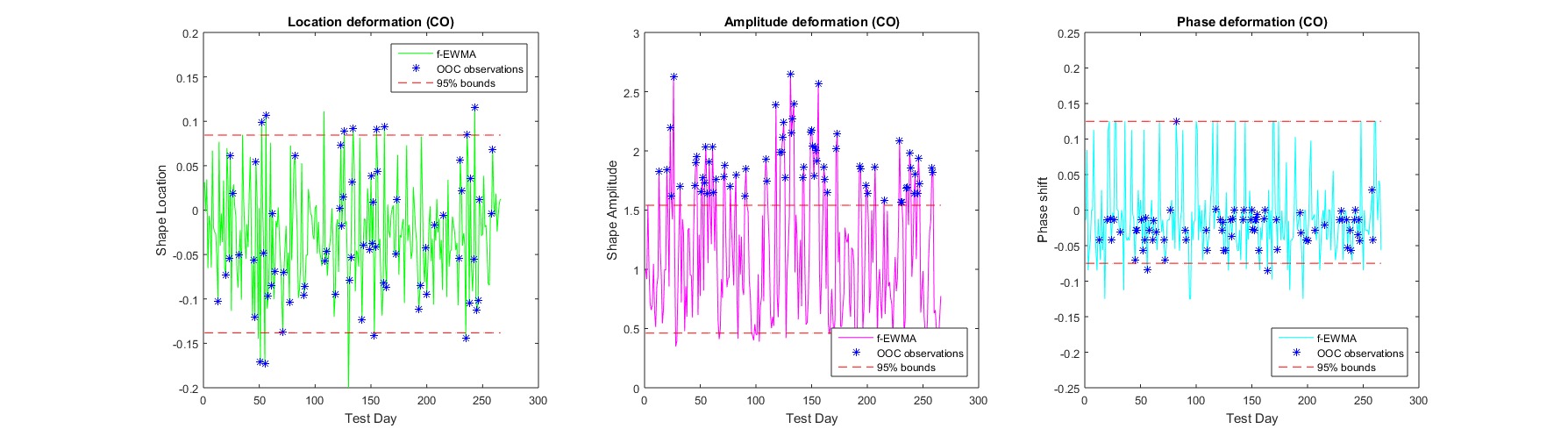}
	\caption{The EWFMA control charts for monitoring shape and deformation deviances (upper panel) and for the individual deformation characteristics (lower panel) for CO.}\label{fig-4.4}
\end{figure}

\begin{figure}[ht!]
	\centering
	\includegraphics[width=4.2in]{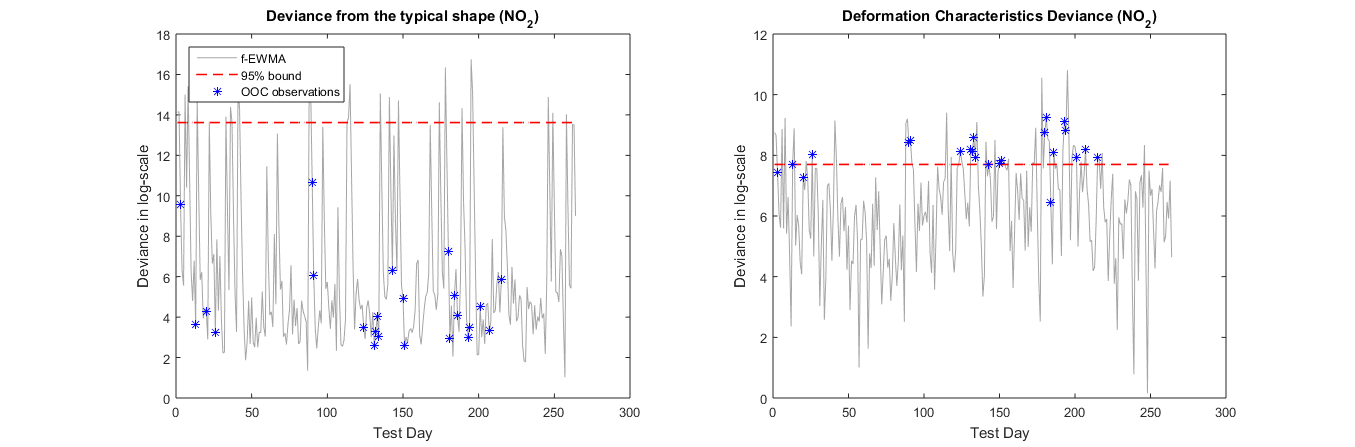}\\
	\includegraphics[width=5.7in]{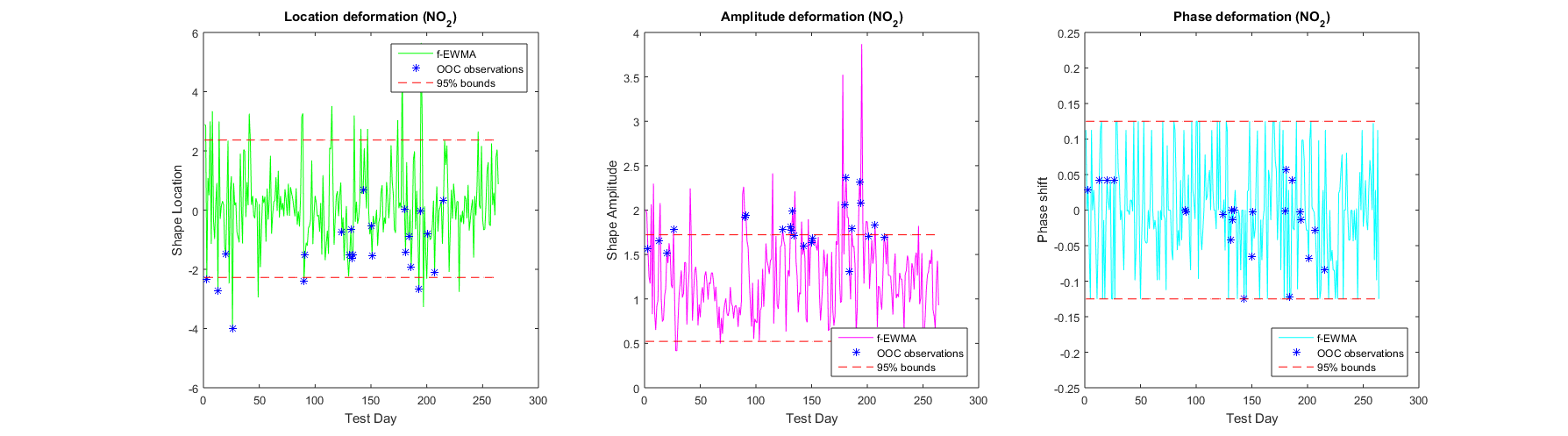}
	\caption{The EWFMA control charts for monitoring shape and deformation deviances (upper panel) and for the individual deformation characteristics (lower panel) for NO$_2$.}\label{fig-4.5}
\end{figure}

\begin{figure}[ht!]
	\centering
	\includegraphics[width=4.2in]{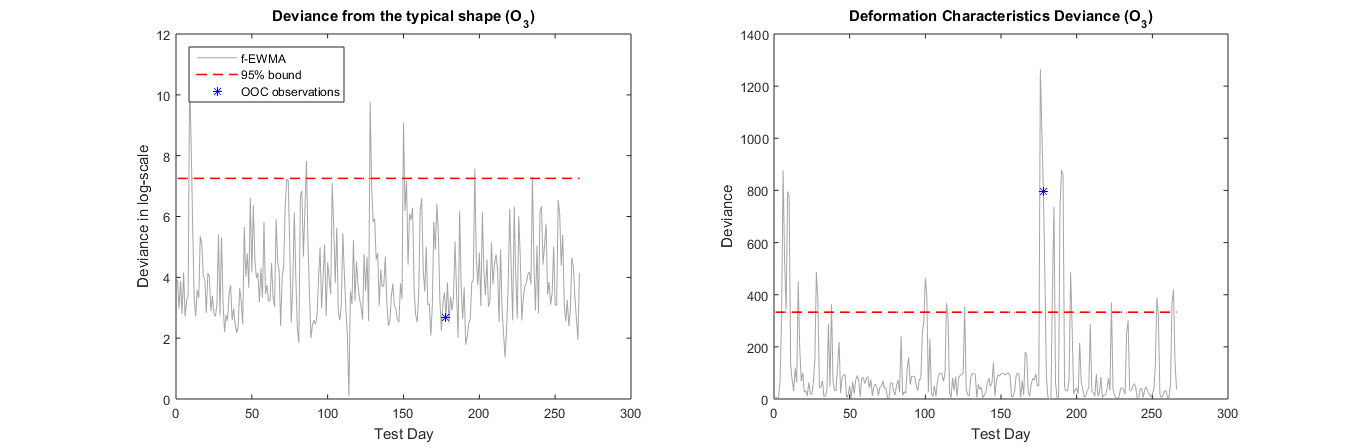}\\
	\includegraphics[width=5.7in]{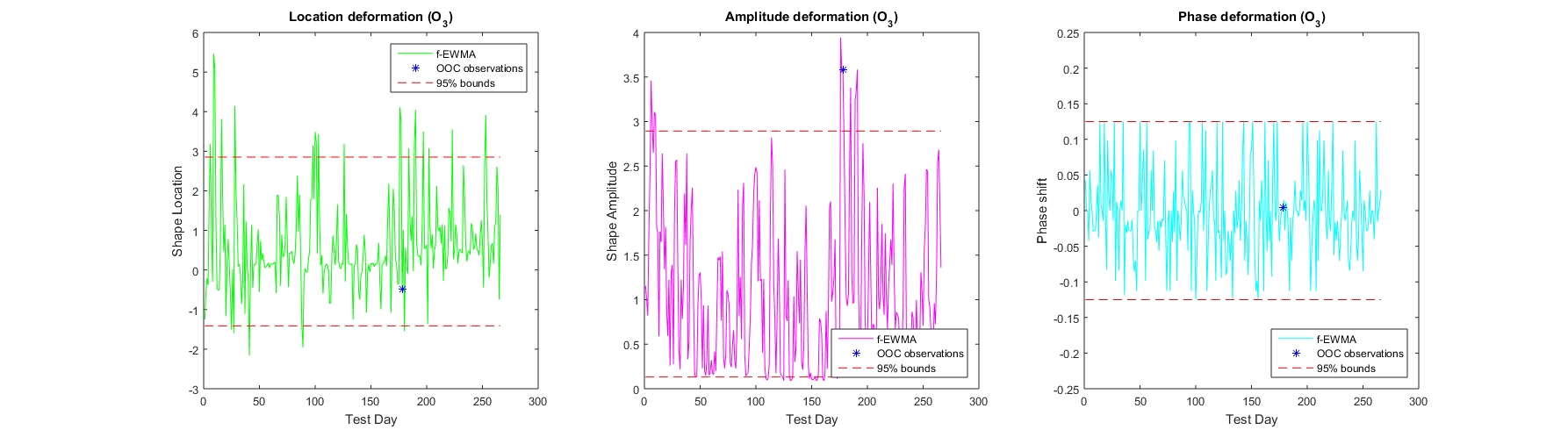}
	\caption{The EWFMA control charts for monitoring shape and deformation deviances (upper panel) and for the individual deformation characteristics (lower panel) for O$_3$}\label{fig-4.6}
\end{figure}

\begin{figure}[ht!]
	\centering
	\includegraphics[width=4.2in]{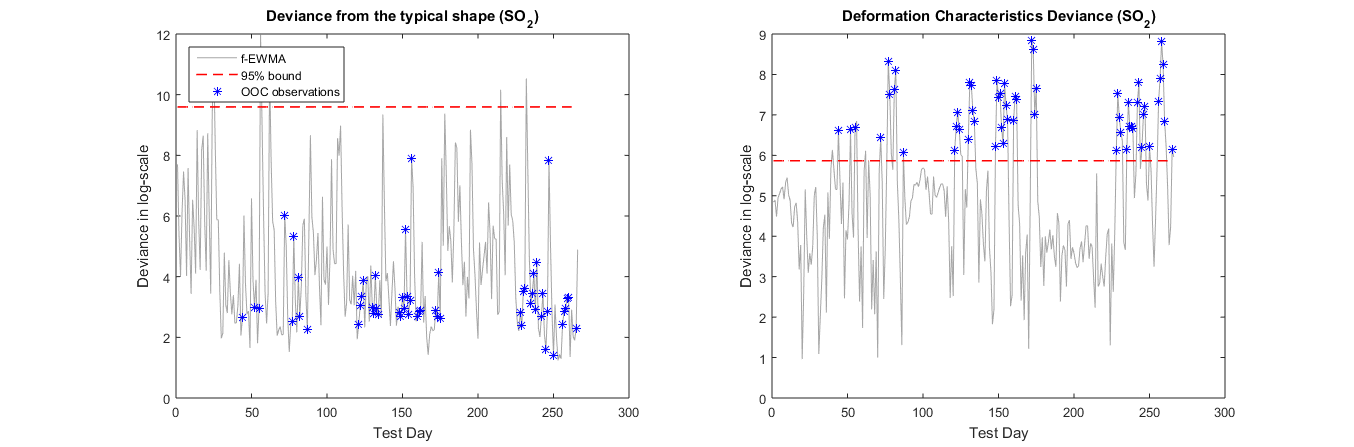}\\
	\includegraphics[width=5.7in]{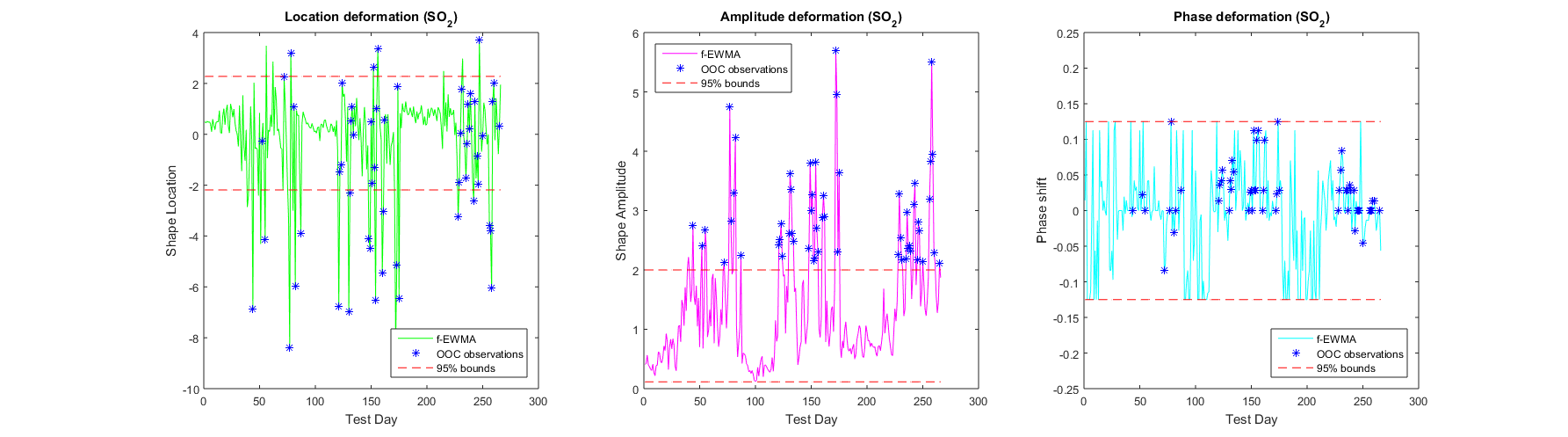}
	\caption{The EWFMA control charts for monitoring shape and deformation deviances (upper panel) and for the individual deformation characteristics (lower panel) for SO$_2$}\label{fig-4.7}
\end{figure}

Subsequently, the phase II of the functional monitoring scheme is performed and the method is assessed over the test dataset concerning the time period 2005-2007. Following the described procedure, the EWFMA control charts for the sampled profiles are constructed to detect significant shifts either in shape and/or in the deformation process. The resulting diagnostic charts are depicted for each pollutant under study in Figures \ref{fig-4.4}, \ref{fig-4.5}, \ref{fig-4.6} and \ref{fig-4.7}. In the upper panel, the deviance related charts are displayed with respect to the reference shape (left column) and the total deformation effect with respect to the estimated mean pattern (right column). The second panel of charts concerns shifts with respect to the individual deformation features explained by the shape-invariant model (shape location, shape amplitude and phase shift). Note that the Fr\'echet mean is calculated according to the $L^2$-metric sense, while exponential weighting parameters on the grid 
$$\lambda=0.01, \,\, 0.02,\,\, \ldots ,\,\, 0.05,\,\, 0.10,\,\, \ldots,\,\, 0.90,\,\, 0.95,\,\, 0.96,\,\, \ldots, \,\, 0.99$$
have been tested. The optimal choice was selected per case (the term optimality refers to the choice of $\lambda$ for which less OOC observations are classified as IC and less IC observations are classified as OOC, i.e. minimizing both type-I and type-II errors). Although, one could try to occasionally reallocate the weighting parameter value by separating the train set to further parts and re-arranging the values according to cross-validation findings, such a task is beyond the scopes of this work and therefore a constant value for each pollutant for the whole monitoring task is used.

\begin{table}[ht!]\scriptsize
	\centering
	\begin{tabular}{c|l|c|rr|rr}
		\hline\hline
		& \bf Monitoring & \bf Number & \multicolumn{2}{c|}{\bf Classification Accuracy} &  \multicolumn{2}{c}{\bf Within Class Accuracy}\\
		\bf Pollutant &\bf Process  & \bf of Cases & \bf True (\%) & \bf False (\%) & \bf OOC (\%)  & \bf IC (\%) \\
		\hline
		& both & 266   & 87.22 (232)    &   12.78 (34)   & 100.00 (71)      &  82.56 (161)\\
		CO & shape process               & 266   & 69.55 (185)    &  30.45 (81)    &       2.82$\,\,\,$ (2)      &  93.85 (183)\\
		& deformation process   & 266   &  91.35 (243)    &    8.65 (23)    & 100.00 (71)      &  88.21 (172)\\
		\hline
		
		& both & 264   & 82.58 (218)   &  17.42 (46)  &   78.26 (18)   &  82.99 (200)\\ 
		NO$_2$ & shape process                & 264   & 81.82 (216)   &  18.18 (48)   &   0.00$\,\,$ (0)      &  89.63 (216)\\ 
		& deformation process     & 264   & 84.85 (224)  &  15.15 (40)   &   78.26 (18)  &  85.48 (206)\\ 
		\hline
		
		&  both                      & 266   & 88.35 (235)  &  11.65 (31)   &  100.00 (1)      &  88.30 (234)\\ 
		O$_3$  & shape process            & 266   & 96.99 (258)  &   3.01$\,\,$ (8)     &      0.00$\,$ (0)      &  97.36 (258)\\ 
		& deformation process & 266   & 90.23 (240)  &   9.77 (26)   &  100.00 (1)      &  90.19 (239)\\ 
		\hline
		
		&  both  & 266   & 95.86 (255) &     4.14 (11)  &  100.00 (55)    &  94.79 (200)\\ 
		SO$_2$ &  shape process             & 266   & 77.44 (206) &  22.56 (60) &  0.00$\,\,\,$ (0)    &  97.63 (206)\\ 
		&  deformation process & 266   & 96.62 (257) &    3.38$\,\,\,$ (9)   &  100.00 (55)  &  95.73 (202)\\ 
		
		\hline\hline
	\end{tabular}
	\caption{Per stage performance of the functional EWMA-type monitoring scheme for each pollutant }\label{tab-4.2}
\end{table}

In Table \ref{tab-4.2} are illustrated the performance diagnostics for each monitoring stage of the method concerning: (a) the shape process, (b) the deformation process and (c) the profile in total, relying on the modelling capabilities and explainability of the shape invariant model. The classification accuracy of the scheme is assessed both in total and within each profile category (IC and OOC ones) through the relevant true classification and wrong classification percentages. The Classification Accuracy section of the table refers to the total true classification percentages, i.e. IC and OOC profiles that identified correctly by the scheme, and false classification percentages, i.e. cases of IC profiles that identified as OOC and the opposite. The Within Class Accuracy section, refers to the percentages of correct identifications within each class, i.e. OOC column depicts the percentage of OOC cases that are correctly identified as OOC comparing to the total number of the OOC cases, while IC column depicts the percentage of IC cases that are correctly identified as IC comparing to the total number of the IC cases. The shift detection scheme displayed in general a nice performance in identifying the shifts (according to the threshold definition of WHO discussed above) with true classification percentages between 82--96\%. Clearly, the pollutants CO and NO$_2$ can be characterized as the more difficult cases where higher error rates are observed, while the SO$_2$ is the pollutant where the scheme displayed the best accuracy. From the results, concerning the underlying shape process monitoring (rows about the shape process in Table \ref{tab-4.2} and the first deviance plot in the upper panel of the individual figures per pollutant) it is evident that the SIM is appropriate for describing such type of profiles, since too few cases are indicated as not explainable by this model (OOC cases). Note that in the table the results have been calculated according to the classification by WHO and not according to actual shape shifts identifications. The deformation process shift detection scheme performed also decently in identifying actual shifts for the pollutants CO, O$_3$ and SO$_2$ with total true classification percentage between 90--96\% and zero probabilities in conducting type-I error. However, in the case of NO$_2$ the error rates are much higher (around 15--20\%). The individual deformation characteristics can be further investigated through the individual deformation parameter charts when a shift is detected. For all pollutants, the phase shift attribute quantified by the parameter $\zeta$ seems not to be related to the actual shifts and in the majority of cases, the related charts do not display behaviours beyond the limits. On the contrary, the phase amplitude parameter $\alpha$ and in some cases the vertical shift parameter $\beta$ identify the significant changes (shape deformations) in the daily concentration profiles as displayed from the relevant plots (see second panel of graphs in Figures \ref{fig-4.4},  \ref{fig-4.5},  \ref{fig-4.6} and \ref{fig-4.7}).  In general, it seems that amplitude deformation level can be used as a proxy for detecting significant shifts when shape process is in control.

\subsection{Methodological Overview}\label{sec-4.5}

The presented methodology provides a complete framework for the both tasks of statistical modelling and monitoring of functional profiles, building on the combination of the general applicability of the Fr\'echet Mean (FM) and the explanatory capabilities of the Shape Invariant Model (SIM). The obtained methodological contributions are summarized below: 

\begin{itemize}
\item A compatible version of the FM with the SIM is determined in \eqref{fm-sim}, expressing the typical profile in a semi-parametric form that depends both on (a) the sampled of functional profiles and (b) the deformation features incorporated by SIM. The typical profile (i.e. the respective Fr\'echet mean of the sampled profiles) is determined by the solution of problem \eqref{opt-2}. Since the optimal solution of the problem depends on the nature of the sampled profiles, the regularized version of the problem stated in \eqref{reg-ff} is studied providing a more stable estimation procedure, while the relevant numerical approximation scheme is described in Algorithm \ref{alg-2}. \\

\item The optimal deformation features of the profiles under study with respect to the typical one (estimated on the previous step) are obtained through the solution of the profile registration problem \eqref{profile-reg} for each one of the profiles. This results to a dataset of deformation parameters that allows for specifying deviance of each profile from the mean pattern with respect to the different deformation features that are taken into account by the SIM.\\

\item A modified version of the well known EWMA monitoring scheme is developed that detects potential shifts of newly sampled profiles either on shape, or on the deformation features (or both). The SIM-compatible version of the Fr\'echet mean and the optimal deformation features of the profiles in the training set obtained in the above steps, are employed to characterize the IC behaviour. Potential shifts in the underlying shape of the newly sampled profile are detected by the scheme stated in \eqref{shape-dev}, while potential shifts in its deformation comparing to the typical shape are detected by the scheme stated in \eqref{deformation-dev} which relies to the solution of the optimization problem \eqref{ewma-opt-1}. These steps are performed successively to characterize a newly sampled profile as an IC one or as an OOC one, according to the stages of Algorithm \ref{alg-1} and the illustration provided in Figure \ref{ms}.

\end{itemize}

\section{Conclusions}\label{sec-5}

In this work, an appropriate framework for the statistical treatment and monitoring of functional profiles is presented. Relying on the notion of the Fr\'echet mean (which offers a generalized sense of the mean for metric space valued data, such as curves), combined with the framework of the the SIM, a modelling procedure has been proposed that allows both for (a) the characterization of the typical (mean) profile and (b) the description of acceptable deformations/deviations from the typical profile in terms of the deformation features that are incorporated by the SIM (i.e. amplitude and phase deformations). At the next step, the obtained typical pattern (i.e. the obtained Fr\'echet mean of the set of profiles under study) and the induced sample of deformation parameters and their distributions are used to develop a monitoring scheme compatible to the functional data setting. The proposed scheme is an extension, for functional data, of the well known EWMA monitoring process, employing the notion of the the Fr\'echet mean combined with the Shape Invariant Model (SIM). The scheme proceeds along two sequential stages. First, potential shifts (deformations) of the underlying shape of the profile under study are examined as compared to the typical shape. If a shift is detected at this stage the sampled profile is considered as an OOC one (shape shift). In case that no shift is detected in the shape process, the second stage of the scheme examines if the deformations of the profile lie within the acceptable standards of deviations from the typical behaviour. Potential shifts in the deformation process can be further allocated to certain deformation features like amplitude deformations, phase deformations, etc.

For the proper illustration and assessment of the presented methodology, air pollution profiles for pollutants from an area in the city of Athens were studied. The described modelling framework has been implemented to capture the typical intra-day patterns of the pollutants and the relevant distribution of the deformation features that characterize the acceptable deviance from the IC behaviour. Next, the developed monitoring scheme was implemented for detecting shifts in the daily sampled profiles of the pollutants with success. In particular, the total accuracy in identifying correctly the status of the profiles varies between 83\% and 96\%. For the IC category, the percentage of true identification varies between 83\% and 95\%. Moreover, concerning the OOC category, for the three pollutants all OOC profiles were successfully identified with the exception of one pollutant where the percentage of true classification was about 78\%. Although the percentages of correct identification are quite high in most of the cases, the monitoring scheme seems to present a more conservative behaviour when characterizing profiles as IC leading to higher Type II errors, while Type I errors are almost zero. The higher Type II errors in these cases could be reduced by including more information in the monitoring task (e.g. take into account extra environmental quantities like temperature, humidity, etc), or by more careful modelling of the interdependencies concerning the deformation features within and across the different pollutants, or even by considering a different type of deformation model to SIM in case that the under study pollutant may provide quite different patterns which should be still considered as IC ones (e.g. by using landmark deformation models \cite{gervini2004self}). In any case, all the above directions can be considered as extensions of the current framework and potential steps for future research.

\subsubsection*{Acknowledgements:} The authors would like to thank the referees for their comments and suggestions that greatly improved the quality of presentation of the paper. 

\subsubsection*{Funding:} Two of the authors wish to acknowledge for financial support from the research program DRASI II, funded by the AUEB Research Center.

\subsubsection*{Data Availability:} Data used in this paper are publicly available from the official link: \url{https://ypen.gov.gr/perivallon/poiotita-tis-atmosfairas/dedomena-metriseon-atmosfairikis-rypansis/}

\subsubsection*{Abbreviations:} The following abbreviations are used in this manuscript:\\

\noindent 
\begin{tabular}{ll}
EWFMA & Exponentially Weighted Fr\'echet Moving Average\\
EWMA & Exponentially Weighted Moving Average\\
IC  & In control\\
OOC & Out of control\\
PCA & Principal Components Analysis\\
SIM & Shape Invariant Model\\
WHO & World Health Organization
\end{tabular}

\appendix

\section{Proof of Main Results in Section \ref{sec-3}}\label{App-A}

\subsection{Proof of Proposition \ref{prop-1}}\label{App-A-1}

\begin{proof}[Proof of Proposition \ref{prop-1}]
	Assume that $f_0$ and $f_j$ for $j=1,2,...,n$ are sufficiently smooth functions. The objective function of the regularized problem, $J_{\rho}(\btheta)$, is a continuous functions since both $V$ and the regularization terms are continuous and hence, by standard arguments, it admits a solution in the compact set 
	$$ \Theta := \left\{ \btheta=(\balpha', \bbeta', \bkappa', \bzeta')' \in \R^{4n}\,\, : \,\, \mbox{satisfying \eqref{c-cons}} \mbox{ and } \begin{array}{l} \alpha_j \in (0, \alpha_{\max}] \\ \beta_j \in [\beta_{\min}, \beta_{\max}] \\
	\kappa_j \in (0, \kappa_{\max}] \\ \zeta_j \in [ \zeta_{\min}, \zeta_{\max} ] \end{array}, \,\, \forall j=1,2,...,n \right\}$$
	where the lower and upper bounds are specified subject to the provided profiles $f_0$ and $f_j$ for $j=1,2,...,n$. Moreover, note that the regularization term $R_{\rho}(\btheta) = \frac{1}{2\rho}|\btheta\|_2^2$ for any $\rho>0$ is a strictly convex function. By continuity arguments on the derivative of $J_{\rho}$ for sufficiently small $\rho$, $J_{\rho}$ remains strictly convex and hence admits a unique minimum. The critical value $\rho_0$ depends on the magnitude of the elements of the Hessian matrix of $V$. 
\end{proof}

\subsection{Proof of Proposition \ref{prop-2}}\label{App-A-2}

\begin{proof}[Proof of Proposition \ref{prop-2}]\mbox{}\\ 
	(a). Fixing $\bxi^{(0)} = \begin{pmatrix} \bkappa^{(0)} \\ \bzeta^{(0)} \end{pmatrix} \in \Theta_{\Pi}$, the regularized Fr\'echet function with respect to $\bgamma = \begin{pmatrix} \bara\\ \barb \end{pmatrix}$ and for a certain choice $\rho>0$ becomes
	\begin{eqnarray}\label{fun-1}
		J_{\rho}(\bgamma) &:=& V(\bgamma, \bxi^{(0)}) + \frac{1}{2\rho} ( \|\bara\|_2^2 + \|\barb\|_2^2 + \| {\bm \xi}^{(0)} \|_2^2)\\
		&=& \frac{1}{n} \bara^T \widetilde{C} \bara + \frac{1}{n} \barb {\bf 1} {\bf 1}^T \barb + {\bf 1}_n^T \hat{F} \nonumber \\
		&& - \frac{2}{n}  {\bf 1}_n^T \widetilde{G} \bara - \frac{2}{n}  {\bf 1}_n^T \barb F^T {\bf 1}_n + \frac{2}{n}  {\bf 1}_n^T \barb \tF^{T} \bara + \frac{1}{2\rho}\left( \bara^{T}\bara + \barb^T\barb + ({\bm \xi}^{(0)})^T {\bm \xi}^{(0)} \right) \nonumber
	\end{eqnarray}
	where 	$\widetilde{f}_j(t) := f_j( \kappa^{(0)}_j t + \zeta^{(0)}_j )$ for $j=1,2,...,n$ and
\begin{eqnarray*}
	F := \begin{pmatrix} \int_I f_1(s) ds\\ \int_I f_2(s) ds\\ \vdots \\ \int_I f_n(s) ds \end{pmatrix} \in \R^n, \,\,\,\,
	\widetilde{F} := \begin{pmatrix} \int_I f_1(\kappa^{(0)}_1 s + \zeta^{(0)}_1) ds\\ \int_I f_2(\kappa^{(0)}_2 s + \zeta^{(0)}_2) ds\\ \vdots \\ \int_I f_n(\kappa^{(0)}_n s + \zeta^{(0)}_n) ds \end{pmatrix} \in \R^n, \,\,\,\,
	\hat{F} := \begin{pmatrix} \langle f_1, f_1 \rangle \\ \langle f_2, f_2 \rangle\\ \vdots \\ \langle f_n, f_n \rangle\ \end{pmatrix} \in \R^n, \\
	\widetilde{C} := 
		\begin{pmatrix} 
			\langle \widetilde{f}_1, \widetilde{f}_1 \rangle & 
			\langle \widetilde{f}_1, \widetilde{f}_2 \rangle &
			\ldots & 
			\langle \widetilde{f}_1, \widetilde{f}_n \rangle \\
			\langle \widetilde{f}_2, \widetilde{f}_1 \rangle & 
			\langle \widetilde{f}_2, \widetilde{f}_2 \rangle &
			\ldots & 
			\langle \widetilde{f}_2, \widetilde{f}_n \rangle \\
			\vdots & \vdots & \ddots & \vdots\\
			\langle \widetilde{f}_n, \widetilde{f}_1 \rangle & 
			\langle \widetilde{f}_n, \widetilde{f}_2 \rangle &
			\ldots & 
			\langle \widetilde{f}_n, \widetilde{f}_n \rangle 
		\end{pmatrix} \in \R^{n \times n}, \,\,\,\, 
	\widetilde{G} := 
		\begin{pmatrix} 
			\langle f_1, \widetilde{f}_1 \rangle & 
			\langle f_1, \widetilde{f}_2 \rangle &
			\ldots & 
			\langle f_1, \widetilde{f}_n \rangle \\
			\langle f_2, \widetilde{f}_1 \rangle & 
			\langle f_2, \widetilde{f}_2 \rangle &
			\ldots & 
			\langle f_2, \widetilde{f}_n \rangle \\
			\vdots & \vdots & \ddots & \vdots\\
			\langle f_n, \widetilde{f}_1 \rangle & 
			\langle f_n, \widetilde{f}_2 \rangle &
			\ldots & 
			\langle f_n, \widetilde{f}_n \rangle 
		\end{pmatrix} \in \R^{n \times n}.
\end{eqnarray*}

Since function \eqref{fun-1} is quadratic, taking first order conditions with respect to $\bara$ and $\barb$, the linear system is obtained
\begin{equation}\label{ff-lin-system}
	\begin{pmatrix}
		\widetilde{C} + \frac{1}{\rho}{\bm I} & \tF {\bf 1}^{T}_n\\
		{\bf 1}_n \tF^{T} & {\bf 1}_n {\bf 1}_n^T + \frac{1}{\rho}{\bm I}
	\end{pmatrix} 
	\begin{pmatrix}
		\bara \\ \barb
	\end{pmatrix} = 
	\begin{pmatrix}
		\widetilde{G}^T {\bf 1}_n \\
		{\bf 1}_n {\bf 1}_n^T F
	\end{pmatrix}
\end{equation}
where ${\bm I}$ denotes the identity matrix in $\R^{n \times n}$. Then, there exists a maximum value $\rho^*_{\Lambda}$ for which the linear system \eqref{ff-lin-system} is explicitly solved for any $\rho \in (0, \rho^*_{\Lambda}]$ (since $J_{\rho}$ becomes strictly convex). 

Let us fix $\bgamma^{(0)} = \begin{pmatrix} \bara^{(0)}\\ \barb^{(0)} \end{pmatrix} \in \Theta_{\Lambda}$. The regularized Fr\'echet function with respect to $\bxi$, 
$$ J_{\rho}( \bxi ) := V(\bgamma^{(0)}, \bxi) + \frac{1}{2\rho}( \|\bxi\|_2^2 + \|\bgamma^{(0)}\|_2^2 )$$ is a continuous function since both $V$ and the regularization terms are continuous and hence, by standard arguments, it admits a solution in the compact set $\Theta_{\Pi}$. Since, for any $\rho>0$ the regularization term is a strictly convex function, by continuity arguments on the derivative of $J_{\rho}$ for sufficiently small $\rho$, $J_{\rho}$ remains strictly convex and hence admits a unique minimum. The critical value $\rho_{\Pi}^*$ depends on the magnitude of the elements of the Hessian matrix of $V$. 

Therefore, defining $\rho^* := \min(\rho^*_{\Lambda}, \rho^*_{\Pi})$, then both problems \eqref{sub-opt-1} and \eqref{sub-opt-2} admit unique solutions for any $\rho$, strictly positive and at most equal to $\rho^*$. Moreover, since the projection operators $Proj_{\Theta_{\Lambda}}$ and $Proj_{\Theta_{\Pi}}$ are single-valued, the solution to the problems \eqref{sub-opt-1} and \eqref{sub-opt-2} are unique. \\ 

\noindent (b). Given that $\rho \in (0,\rho^{*}]$, uniqueness of solutions for the optimization problems \eqref{sub-opt-1} and \eqref{sub-opt-2} are guaranteed at each step of the algorithm by (a). At each iteration, the described numerical splitting scheme, generates a sequence of points $\{ \btheta_k \}_k$ by alternating projections between the two convex sets $\Theta_{\Lambda}$ and $\Theta_{\Pi}$. In particular, at Step 1 the problem is solved on the convex set $\Gamma := \Theta \times \R^{2n}$ generating a sequence of points $\{ \bar{\btheta}_k \}_k \subset \Gamma$, while at Step 2, the problem is solved on the convex set $\Xi := \R^{2n}\times \Theta_{\Pi}$ generating a sequence of points $\{ \tilde{\btheta}_k \}_k \subset \Xi$. It can be shown by the alternating projections algorithm (see e.g. \cite{bauschke1993convergence, bauschke1994dykstra}) that as $k$ grows, both sequences converge to a point $\btheta_*$ in the intersection of the two sets $\mathcal{D} = \Gamma \cap \Xi$, being the minimizer of $J(\btheta) := V(\btheta) + \frac{1}{2\rho}\|\btheta\|_2^2$. 
\end{proof}

\subsection{Proof of Proposition \ref{prop-3} and auxiliary results}\label{App-A-3}

\begin{proof}[Proof of Proposition \ref{prop-3}]
The single profile regularized registration problem admits solutions since the objective function $J_{\rho}^{(j)}(\theta) := V_j(\theta) + \frac{1}{2\rho}\|\theta\|_2^2$ is a continuous function since both $V$ and the regularization terms are continuous and hence, by standard arguments, it admits a solution in the compact set $\Theta_R$ as stated in \eqref{Theta-Reg}. Due to strict convexity of the regularization term for any $\rho>0$ and by continuity arguments on the derivative of $J^{(j)}_{\rho}$ for sufficiently small $\rho$, $J^{(j)}_{\rho}$ remains strictly convex and hence admits a unique minimum. The critical value $\rho_*$ depends on the magnitude of the elements of the Hessian matrix of $V_j$. 
\end{proof}

Moreover, the solution can be parameterized since the amplitude deformation parameters $(\alpha, \beta)$ can be expressed in closed form as functions of the phase deformation parameters $(\kappa,\zeta)$. The result is stated in the following Lemma.

\begin{lemma}\label{lemma-1}
	Given a pair of $(\kappa, \zeta)$ the optimal amplitude deformation parameters $(\alpha, \beta)$ of the single profile registration problem \eqref{sim-reg} admit the closed form
	\begin{eqnarray}
		\alpha_*(\kappa, \zeta) &=& \frac{\sigma_{0,j}(\kappa,\zeta)}{\sigma^2_{0}(\kappa,\zeta)}  \\
		\beta_*(\kappa, \zeta) &=& m_j - \alpha_*(\kappa, \zeta) \,\, m_0(\kappa,\zeta) 
	\end{eqnarray}
	where $f_{0,\kappa,\zeta}(t):=f_0( \kappa^{-1}(t-\zeta))$ denotes the phase deformated mean profile and
	\begin{eqnarray*}
		\begin{array}{ll}
			m_0(\kappa,\zeta) = \int_I f_{0,\kappa,\zeta}(t) dt, & m_j := \int_I f_j(t)dt\\ 
			\sigma_0^2(\kappa,\zeta) := \langle f_{0,\kappa,\zeta}, f_{0,\kappa,\zeta} \rangle - m_0^2(\kappa,\zeta), &
			\sigma_{0,j}(\kappa,\zeta) := \langle f_{0,\kappa,\zeta}, f_j \rangle - m_0(\kappa,\zeta) m_j.
		\end{array}
	\end{eqnarray*}
	the related moment terms.
\end{lemma}

\begin{proof}
	The result is obtained by taking first order conditions to $V_j$ as stated in \eqref{sim-reg} with respect to $\alpha, \beta$ and solving the resulting pair of equations. Continuity and strict convexity of $V_j$ with respect to these parameters guarantee uniqueness for any pair $(\kappa, \zeta)$. 
\end{proof}

\section{Proof of Main Result in Section \ref{sec-4}}
\subsection{Proof of Proposition \ref{prop-4} and auxiliary results}\label{App-B-1}

\begin{proof}[Proof of Proposition \ref{prop-4}]
	The chart element regularized estimation problem admits solutions since the objective function $\tilde{J}_{\rho}^{(j)}(\theta) := V_{\lambda}(\theta) + \frac{1}{2\rho}\|\theta\|_2^2$ is a continuous function since both $V_{\lambda}$ and the regularization terms are continuous and hence, by standard arguments, it admits a solution in the compact set $\Theta_R$ as stated in \eqref{Theta-Reg}. Due to strict convexity of the regularization term for any $\rho>0$ and by continuity arguments on the derivative of $\tilde{J}^{(j)}_{\rho}$ for sufficiently small $\rho$, $\tilde{J}^{(j)}_{\rho}$ remains strictly convex and hence admits a unique minimum. The critical value $\widetilde{\rho}_*$ depends on the magnitude of the elements of the Hessian matrix of $V_{\lambda}$. 
\end{proof}

In the same fashion with Proposition \ref{prop-3}, the solution can be parameterized since the amplitude deformation parameters $(\walpha, \wbeta)$ can be expressed in closed form as functions of the phase deformation parameters $(\wkappa,\wzeta)$. The result is stated in the following Lemma.

\begin{lemma}\label{lemma-2}
	Given a pair of $(\wkappa, \wzeta)$, the optimal amplitude deformation parameters $(\walpha, \wbeta)$ for the chart element estimation problem described in \eqref{ewma-opt-1} admit the closed form
	\begin{eqnarray}
		\widetilde{\alpha}_{\lambda}( \widetilde{\kappa}, \widetilde{\zeta}) &=& \lambda \frac{ \hat{\sigma}_{0,j}(\wkappa, \wzeta) }{ \sigma_0^2(\wkappa, \wzeta) } + (1-\lambda) \frac{ \widetilde{\sigma}_{0,j-1}(\wkappa,\wzeta) }{ \sigma_0^2(\wkappa, \wzeta) } \\
		\widetilde{\beta}_{\lambda}(\wkappa, \wzeta) &=& \lambda \, \hat{m}_j + (1-\lambda) \, \widetilde{m}_{j-1} - \widetilde{\alpha}_{\lambda}(\wkappa, \wzeta) m_0(\wkappa, \wzeta)
	\end{eqnarray}
	where $f_{0,\wkappa,\wzeta}(t):=f_0( \wkappa^{-1}(t-\wzeta))$ denotes the phase deformated typical profile and
	\begin{eqnarray*}
		\begin{array}{ll}
			m_0(\wkappa,\wzeta) = \int_I f_{0,\wkappa,\wzeta}(t) dt, & \sigma_0^2(\wkappa,\wzeta) := \langle f_{0,\wkappa,\wzeta}, f_{0,\wkappa,\wzeta} \rangle - m_0^2(\wkappa,\wzeta)\\
			\hat{m}_j := \int_I \hat{f}_j(t)dt, & \hat{\sigma}_{0,j}(\wkappa,\wzeta) := \langle f_{0,\wkappa,\wzeta}, \hat{f}_j \rangle - m_0(\wkappa,\wzeta) \hat{m}_j \\
			\widetilde{m}_{j-1} := \int_I \widetilde{f}_{j-1}(t)dt, & \widetilde{\sigma}_{0,j-1}(\wkappa,\wzeta) := \langle f_{0,\wkappa,\wzeta}, \widetilde{f}_{j-1} \rangle - m_0(\wkappa,\wzeta) \widetilde{m}_{j-1}.
		\end{array}
	\end{eqnarray*}
	the related moment and variance terms.
\end{lemma}

\begin{proof}
	The result is obtained by taking first order conditions to $V_{\lambda}$ as stated in \eqref{ewma-opt-1} with respect to $\talpha, \tbeta$ and solving the resulting pair of equations. Continuity and strict convexity of $V_{\lambda}$ with respect to these parameters guarantee uniqueness for any pair $(\wkappa, \wzeta)$. 
\end{proof}

\bibliographystyle{chicago}
\bibliography{references_abbr}

\end{document}